\documentclass[english,journal,11pt]{article}
\usepackage[T1]{fontenc}
\usepackage[latin9]{inputenc}
\usepackage{subfigure}
\usepackage{float}
\usepackage{bm}
\usepackage{amsmath}
\usepackage{graphicx}
\usepackage{amssymb}
\usepackage{nomencl}
\makenomenclature
\makeatletter

\floatstyle{ruled}
\newfloat{algorithm}{tbp}{loa}
\floatname{algorithm}{Algorithm}

\newtheorem{definitn}{Definition}
\newtheorem{remrk}{Remark}
\newtheorem{lemma}{Lemma}
\newtheorem{thm}{Theorem}

\newtheorem{prop}{Proposition}

\usepackage{cite}
\usepackage{url}

\makeatother

\usepackage{babel}

\begin{document}

\title{Structured sublinear compressive sensing via belief propagation}
\author{Wei Dai, Olgica Milenkovic and Hoa Vin Pham \\
 Department of Electrical and Computer Engineering, University of Illinois at Urbana-Champaign}

\maketitle

\begin{abstract}
Compressive sensing (CS) is a sampling technique designed for reducing the complexity of sparse 
data acquisition. One of the major obstacles for practical deployment of CS techniques is the signal reconstruction 
time and the high storage cost  of random sensing matrices. 
We propose a new structured compressive sensing scheme, based on codes of graphs, that allows for a
joint design of structured sensing matrices and logarithmic-complexity reconstruction algorithms.
The compressive sensing matrices can be shown to offer asymptotically optimal performance when used in combination
with Orthogonal Matching Pursuit (OMP) methods. For more elaborate greedy reconstruction schemes, we propose a new family of 
list decoding belief propagation algorithms, as well as reinforced- and multiple-basis belief propagation algorithms. Our simulation results indicate
that reinforced BP CS schemes offer very good complexity-performance tradeoffs for very sparse signal vectors.
\end{abstract}
\textbf{Keywords:} Belief Propagation, Compressive sensing, Low-Density Parity-Check codes, Orthogonal Matching
Pursuit, restricted isometry constants, sparse approximation, Subspace Pursuit.

\section{\label{sec:Introduction}Introduction}

Compressive sensing (CS)  has received significant attention
due to their various applications in signal processing, networking, MRI data acquisition, bioinformatics, and 
remote sensing~\cite{Candes_Tao_IT2006_Robust_Uncertainty_Principles}.
CS is a sampling technique for compressible and/or $K$-sparse signals,
i.e., signals that can be represented by $K\ll N$ significant coefficients
over an $N$-dimensional basis. Sampling of a $K$-sparse, discrete-time
signal $\mathbf{x}$ of dimension $N$ is accomplished by computing
a measurement vector, $\textbf{y}$, that consists of $m\ll N$ linear
projections, i.e.,
 \[
\mathbf{y}=\mathbf{\Phi}\mathbf{x}.\]
Here, $\mathbf{\Phi}$ represents an $m\times N$ matrix, usually
over the field of real numbers~\cite{Donoho_IT2006_CompressedSensing}.

Although the reconstruction of the signal $\mathbf{x}\in\mathbb{R}^{N}$
from the possibly noisy random projections is an ill-posed task,
the prior knowledge of signal sparsity allows for recovering
$\mathbf{x}$ in polynomial time using $m\ll N$ observations only.  
If the reconstruction problem is cast as an $\ell_{0}$ minimization problem ~\cite{Bresler_ICASSP1998_sub_Nyquist_sampling}, 
it can be shown that in order to 
reconstruct a $K$-sparse signal $\mathbf{x}$,
$\ell_{0}$ minimization requires only $m=2K$ random projections. In this setting, it is assumed that
the signal and the measurements are noise-free. Unfortunately,
the $\ell_{0}$ optimization problem is a combinatorial problem that for general instances of sensing is NP-hard. 

The work by Donoho and Candes et. al.~\cite{Donoho_IT2006_CompressedSensing,Candes_Tao_IT2005_decoding_linear_programming,
Candes_Tao_IT2006_Robust_Uncertainty_Principles,
Candes_Tao_FOCS05_Error_Correction_Linear_Programming}
demonstrated that CS reconstruction is a polynomial time
problem -- conditioned on the constraint that more than $2K$ measurements
are used. The key idea behind their approach is that it is not necessary to resort to $\ell_{0}$ optimization to recover $\mathbf{x}$
from the under-determined inverse problem: a much easier $\ell_{1}$
optimization, based on Linear Programming (LP) techniques, yields
an equivalent solution provided that the sensing matrix $\mathbf{\Phi}$
satisfies the so called \emph{restricted isometry property} (RIP),
with a constant RIP parameter. 

While LP techniques play an important role in designing computationally
tractable CS decoders, their complexity renders them highly impractical
for many applications. In such cases, the need for fast reconstruction
algorithms -- preferably operating in time linear in $N$, and without significant performance loss compared to LP methods
 -- is of critical importance. A common approach to mitigating these problems is to increase the number of measurements
 and to use greedy reconstruction methods.
Several classes of low-complexity reconstruction techniques were put forward as alternatives to linear programming (LP) recovery,
including group testing methods \cite{Cormode_2006_Combinatorial},
pursuit strategies such as  Orthogonal Matching Pursuit (OMP), Subspace Pursuit (SP) and Compressive Sampling Matching Pursuit (CoSaMP)
~\cite{Tropp_2007_OMP,Wei-Olgica-Subspace-Pursuit,Tropp_ITA2008_Iterative_Recovery,cosamp},
and coding-theoretic techniques~\cite{Baraniuk_2008_Bayesian_CS_BP,donoho,dimakis}.

We focus our attention on two intertwined problems related to low-complexity CS reconstruction techniques. The first problem is concerned
with designing structured matrices that provide RIP-type
performance guarantees, since such matrices have low storage complexity and may potentially yield to faster reconstruction approaches. 
The second problem is concerned with how to most efficiently exploit the structure of the sensing matrix in order to
further reduce the reconstruction complexity of greedy-like methods that use correlation maximization as one of their key steps. 
The solution we propose addresses both issues, and can be succinctly described as follows.

It is known that random Bernoulli matrices -- matrices with i.i.d. Bernoulli(1/2) distributed entries -- have constant RIP parameters with a number of measurements
proportional to $K \log(N/K)$~\cite{Candes_Tao_IT2006_Robust_Uncertainty_Principles,Candes_Tao_FOCS05_Error_Correction_Linear_Programming}. This number 
of measurements suffices for exact reconstruction of $K$-sparse
signals using LP methods. One property of Bernoulli matrices is that, for sufficiently large dimensions,
the fraction of the symbols $+1$ and $-1$ per row and per column is close to one half. Furthermore, a similar property holds for any sufficiently large submatrix
of the matrix. One approach to designing structured compressive sensing matrices would be to try mimicking this property of Bernoulli matrices and then showing 
that the matrices indeed have a constant RIP parameter. 

This task can be accomplished via linear error-correcting coding. Due to the linear structure of the code, using codewords of a binary linear code with zeros replaced 
by $+1$'s and ones by $-1$'s as columns of the matrix ensures the row-weight balancing property. Furthermore, if the weight of the codewords is chosen close to 
half of the codelength, similar concentration results will hold for the columns of the sensing matrix.

The idea of using linear error-correcting codes was first proposed in \cite{calderbank}, where encodings of Reed-Muller codewords were used for columns of
a compressive sensing matrix \cite{haupt-bajwa,calderbank1}. The authors proposed independently a similar framework based on low-density parity-check codes \cite{gallager} in 
\cite{our-isit}, and some follow-up results on this work were reported in~\cite{arya}. 
Another approach for constructing sensing matrices by trying to match their distribution of singular values to that of Bernoulli matrices was put forward in~\cite{Babadi2010_Binary_Random_Matrices,babadi-tarokh}. 

The advantage of using sensing matrices based on error-control codes from the perspective of reconstruction complexity is best explained in the context of greedy algorithms, as argued in
our earlier work \cite{our-isit}. A key step
of greedy reconstruction algorithms is to compute the correlations of the observed vector $\mathbf{y}$ with the columns of the sensing matrix $\mathbf{\Phi}$, and to identify
the column with the largest correlation. When the columns of the matrix represent codewords of a linear code, this problem reduces to the extensively studied maximum likelihood (ML) decoding
problem. For certain classes of codes, near-ML decoding can be performed in time linear in the length of the code, which in the described setting implies that near-optimal correlation optimization 
can be performed in time proportional to the number of rows, and not the number of columns of the sensing matrix.

We focus on codes that lead to reconstruction techniques with
sublinear -- more precisely -- logarithmic complexity in $N$. The basic construction and decoding methods are based on ideas from codes on graphs and iterative
decoding. We show how a simple combination of reinforced belief propagation (BP)~\cite{Braunstein_ReinforcedBP} and a novel list decoding method can be coupled with the greedy SP algorithm to produce good reconstruction
algorithms with logarithmic complexity, for the case of ``super-sparse'' signals previously studied in~\cite{Gilbert_Tropp_CS}. As already mentioned, the BP algorithm operates on the columns of the matrix
$\mathbf{\Phi}$ of length $m$, and consequently, its reconstruction complexity is
$O\left(m\right)$. 

Before outlying the organization of the paper, we would like to describe the context of our work within the vast literature on compressive sensing. Sublinear reconstruction techniques were first investigated in~\cite{Gilbert_Tropp_CS,Sarvotham_CSBP,berinde1,gilbert}, while \emph{sparse sensing matrices} 
coupled with BP decoding were considered in \cite{Baraniuk_2008_Bayesian_CS_BP,berinde1}. An idea for sublinear compressive sensing reconstruction inspired by Sudoku was described in~ \cite{Sarvotham_CSBP},
but the algorithm works only for input signals with special structural properties where one requires that all sums of subsets of coefficients are distinguishable (which is rather restrictive for
binary vectors), and where the measurement matrix is random. Furthermore, the reconstruction is only partial, in so far that the reconstruction complexity strongly depends on the number of recovered entries of the sensed vector. 

Our approach differs from all the aforementioned results in so far that it \emph{does not use sparse sensing matrices} that are known to incur a performance loss compared to dense matrices, such as Bernoulli matrices.
Although our structured sensing matrices are dense, they are constructed using codewords of large minimum distance LDPC codes which themselves have sparse matrix descriptions (i.e., sparse parity-check matrices). 
Furthermore, no high-complexity pre-processing is required and unlike the approach in~\cite{Gilbert_Tropp_CS}, the complexity of the algorithm is not polylog in $N$, 
but only logarithmic in $N$; and, as opposed to using sparse matrices without RIP guarantees, our approach utilizes \emph{structured dense matrices} constructed from sparse matrices that are
asymptotically optimal with respect to the achievable coherence parameters. 

The problems addressed in this paper are equally relevant to questions arising in storage and wireless communication systems, since a major part of analysis is focused on 
BP decoding for channels with severe user interference. The framework proposed in this paper also allows for handling measurement noise, 
but the underlying results will be described elsewhere.

The paper is organized as follows. Section \ref{sec:Compress-Sensing} provides a brief introduction to compressive sensing. 
Section \ref{sec:LDPC_CS} includes the description of a structured design approach for compressive sensing matrices $\mathbf{\Phi}$, amenable for $O\left(K^{u}\log N\right)$ 
complexity decoding of super-sparse vectors, with $u=2$ or $u=3$. This section also contains the main analytical results of the paper. 
Section \ref{sec: Algorithm} describes a new ``biased list decoding'' framework for BP algorithms, a new CS-oriented reinforced BP algorithm, as well as 
the description of multiple-basis belief propagation algorithm for CS reconstruction. Section \ref{sec:simulation} presents the simulation results, while Section \ref{sec:conclusion} contains our concluding remarks.

\section{\label{sec:Compress-Sensing}Compressive Sensing and the Restricted Isometry Property}

Let $\mathbf{x}$ be an $N$-dimensional real-valued signal with at most $K$ non-zero components, henceforth
called a $K$-sparse signal.
Let $\text{supp}(\mathbf{x})$ denote the set of indices of the non-zero
coordinates of the vector $\mathbf{x}=(x_{1},\ldots,x_{N})$,
and let $|\text{supp}(\mathbf{x})|=\|\cdot\|_{0}$
denote the support size of $\mathbf{x}$, or equivalently,
its $\ell_{0}$ norm.
Assume next that $\mathbf{x}\in\mathbb{R}^{N}$ is an unknown signal
with $|\text{supp}(\mathbf{x})|\leq K$, and assume that  $\mathbf{y}\in\mathbb{R}^{m}$ is 
an observation of $\mathbf{x}$ generated via $m$ linear measurements, i.e.,
$\mathbf{y}=\mathbf{\Phi}\mathbf{x}$,
 where $\mathbf{\Phi}\in\mathbb{R}^{m\times N}$ is referred
to as the \emph{sensing matrix}.

We are concerned with the problem of low-complexity recovery of the
unknown signal $\mathbf{x}$ from the measurement $\mathbf{y}$ for the case
that $K << N$. A
natural formulation of the recovery problem is within an $\ell_{0}$
norm minimization framework, which seeks a solution to the problem
\[
\min\;\left\Vert \mathbf{x}\right\Vert _{0}\;\mathrm{subject\; to}\;\mathbf{y}=\mathbf{\Phi}\mathbf{x}.\]
 Unfortunately, the above $\ell_{0}$ minimization problem
is NP-hard, and hence cannot be used for practical applications \cite{Candes_Tao_IT2005_decoding_linear_programming}.

One way to avoid using this computationally intractable formulation
is to consider a $\ell_{1}$ optimization problem,
\[
\min\;\left\Vert \mathbf{x}\right\Vert _{1}\;\mathrm{subject\; to}\;\mathbf{y}=\mathbf{\Phi}\mathbf{x},\]
 where \[
\left\Vert \mathbf{x}\right\Vert _{1}=\sum_{i=1}^{N}\left|x_{i}\right|\]
 denotes the $\ell_{1}$ norm of the vector $\textbf{x}$.

The main advantage of the $\ell_{1}$ minimization approach is that
it is a convex optimization problem that can be solved efficiently
via linear programming (LP) techniques. This method is therefore frequently
referred to as $\ell_{1}$-LP reconstruction \cite{Candes_Tao_IT2005_decoding_linear_programming,Candes_Tao_IT2006_Near_Optimal_Signal_Recovery},
and its reconstruction complexity equals $O\left(m^{2}N^{3/2}\right)$ for small $m$, and $O\left(N^3\right)$ for large $m$. The method 
is based on interior point LP solvers \cite{Nesterov_book1994_Interior_point_Convex_Programming}.

The reconstruction accuracy of the $\ell_{1}$-LP method is described by
the \emph{restricted isometry property} (RIP),
formally defined below.

\begin{definitn}
\label{def:Truncation}Let $\mathbf{\Phi}\in\mathbb{R}^{m\times N}$,
$\mathbf{x}\in\mathbb{R}^{N}$ and $I\subset\left\{ 1,\cdots,N\right\} $.
Also, let the matrix $\mathbf{\Phi}_{I}$ consist of the columns of $\mathbf{\Phi}$
indexed by $i\in I$; similarly, let $\mathbf{x}_{I}$ denote a vector composed of the entries
of $\mathbf{x}$ indexed by the same set $I$. The space spanned by the columns
of $\mathbf{\Phi}_{I}$ is denoted by $\mathrm{span}\left(\mathbf{\Phi}_{I}\right)$.
\end{definitn}

\begin{definitn}
\label{def:RIP}A matrix $\mathbf{\Phi}\in\mathbb{R}^{m\times N}$
is said to satisfy the Restricted Isometry Property (RIP) with parameters
$\left(K,\delta\right)$ for $K\le m$, $0\leq\delta\leq1$, if for
all index sets $I\subset\left\{ 1,\cdots,N\right\} $ such that $\left|I\right|\le K$,
and for all $\mathbf{q}\in\mathbb{R}^{\left|I\right|}$, one has
\begin{equation}
\left(1-\delta\right)\left\Vert \mathbf{q}\right\Vert _{2}^{2}\le\left\Vert \mathbf{\Phi}_{I}\mathbf{q}\right\Vert _{2}^{2}\le\left(1+\delta\right)\left\Vert \mathbf{q}\right\Vert _{2}^{2}.
\label{eq:RIP}
\end{equation}

We define $\delta_{K}$, the RIP constant, as the infimum of all parameters $\delta$
for which the RIP holds, i.e. 
\begin{equation}
\delta_{K} :=\inf\;\left\{ \delta:\;\left(1-\delta\right)\left\Vert \mathbf{q}\right\Vert _{2}^{2}\le\left\Vert \mathbf{\Phi}_{I}\mathbf{q}\right\Vert _{2}^{2}\le\left(1+\delta\right)\left\Vert \mathbf{q}\right\Vert _{2}^{2},\right.\\
 \left.\;\forall\left|I\right|\le K,\;\forall\mathbf{q}\in\mathbb{R}^{\left|I\right|}\right\} .\end{equation}

\end{definitn}


\begin{remrk}
Most known families of matrices satisfying
the RIP property with optimal or near-optimal performance guarantees
are random \cite{Candes_Tao_IT2006_Near_Optimal_Signal_Recovery,baraniuk-jl}. Note that the storage complexity of such random matrices satisfying the RIP property is very large.
Alternative sensing matrix design methods rely on structured designs which may mitigate this problem (see \cite{DeVore_2007_Deterministic_CS_matrix,jafarpour,bourgain,haupt-bajwa,babadi-tarokh,
calderbank1}).
\end{remrk}

There exists an important connection between the LP reconstruction
accuracy and the RIP property, first described by Candés and Tao in
\cite{Candes_Tao_IT2005_decoding_linear_programming}. If the sampling
matrix $\mathbf{\Phi}$ satisfies the RIP with constants $\delta_{K}$,
$\delta_{2K}$, and $\delta_{3K}$, such that \begin{equation}
\delta_{K}+\delta_{2K}+\delta_{3K}<1,\label{eq:three-delta-sum}\end{equation}
 then the $\ell_{1}$-LP algorithm will reconstruct all $K$-sparse
signals exactly. 

An alternative to $\ell_{1}$ methods is the family of greedy algorithms, including OMP, Regularized
OMP, Stagewise OMP, SP and CoSaMP algorithms \cite{Tropp_2007_OMP, Vershynin_ROMP_proposal,
Donoho_submitted_StOMP, Wei-Olgica-Subspace-Pursuit,Tropp_ITA2008_Iterative_Recovery}.
The basic idea behind these methods is to find the support of the unknown signal
sequentially. At each iteration of the algorithms, one or several
coordinates of the vector $\textbf{x}$ are selected for testing, based
on the correlation magnitudes between the columns of $\mathbf{\Phi}$
and the regularized measurement vector. If deemed sufficiently reliable,
the candidate column indices are used for the current estimate of the
support set of $\mathbf{x}$. The greedy algorithms iterate this
procedure until all the coordinates in the correct support set are included in
the estimated support set or until reconstruction failure is declared. 

\begin{table}[h] 
\label{tab:omp}
\caption{The OMP Algorithm}

\noindent \textbf{Input}: $\mathbf{y}$, $\mathbf{\Phi}$ and an error threshold $\epsilon_0$

\noindent \textbf{Initialization}: Initialize $k=0$ and set:

\begin{itemize}
\item The initial solution to $\mathbf{x}^0=0$.
\item The residual, $\mathrm{resid}\left(\mathbf{y},\mathbf{\Phi}\right)$, to $\mathbf{r}^0=\mathbf{y}-\mathbf{\Phi}\mathbf{x}^0=\mathbf{y}$.
\item The initial support set to $\mathcal{S}^0=\text{supp}\left(\mathbf{x}^0\right)=\varnothing$.
\end{itemize}
\noindent \textbf{Main iteration}: Increase $k$ by 1 and perform the following steps

\begin{itemize}
\item Compute the errors $\epsilon(j)=\min_{z_j}\;\left\Vert z_j\;\boldsymbol{\varphi}_j-\mathbf{r}^{k-1}\right\Vert _{2}^{2}$, where $\boldsymbol{\varphi}_j$ denotes the 
$j$-th column of $\mathbf{\Phi}$ and $z_j \in \mathbb{R}$.
\item Update the support: Find a ``minimizer'', $j_0$, of $\epsilon(j):   j \notin \mathcal{S}^{k-1},\epsilon(j_0)\leq\epsilon(j)$, and update $\mathcal{S}^k=\mathcal{S}^{k-1}\cup {j_0}$.
\item Compute $\mathbf{x}^k$, the ``minimizer'' of $\left\Vert \mathbf{\Phi x}-\mathbf{y}\right\Vert _{2}^{2}$
subject to $\text{supp}\left(\mathbf{x}\right)=\mathcal{S}^k$.
\item Update residual: Compute $\mathbf{r}^k=\mathbf{y}-\mathbf{\Phi} \mathbf{x}^k$.
\item Stopping rule: If $\left\Vert\mathbf{r}^k\right\Vert_{2}<\epsilon_0$, stop. Otherwise, run another iteration.
\end{itemize}

\textbf{Output}: The solution is $\mathbf{x}^K$, obtained after $K$ iterations.
\end{table}

\begin{table}[h]  
\label{tab:sp}
\caption{\label{alg:Subspace-Pursuit-Algorithm}The SP Algorithm}
\textbf{Input}: $K$, $\mathbf{\Phi}$, $\mathbf{y}$

\noindent \textbf{Initialization}:

\begin{enumerate}
\item $T^{0}=\left\{ K^{\phantom{*}}\right.$indices corresponding to the
largest magnitude entries in the vector $\left.\mathbf{\Phi}^{*}\mathbf{y}\right\} $, where $\mathbf{\Phi}^{*}$ denotes the transpose of $\mathbf{\Phi}$.
\item $\mathbf{y}_{r}^{0}=\mathrm{resid}\left(\mathbf{y},\mathbf{\Phi}_{T^{0}}\right)$.
\end{enumerate}
\noindent \textbf{Iteration}: At the $\ell^{\mathrm{th}}$ iteration, go through
the following steps

\begin{enumerate}
\item $\tilde{T}^{\ell}=T^{\ell-1}\bigcup$$\left\{ K^{\phantom{*}}\right.$indices
corresponding to the largest magnitude entries in the vector $\left.\mathbf{\Phi}^{*}\mathbf{y}_{r}^{\ell-1}\right\} $.
\item Set $\mathbf{x}_{p}=\mathbf{\Phi}_{\tilde{T}^{\ell}}^{\dagger}\mathbf{y}$.
\item $T^{\ell}=\left\{ K^{\phantom{*}}\right.$indices corresponding to
the largest elements of $\left.\mathbf{x}_{p}\right\} $.
\item $\mathbf{y}_{r}^{\ell}=\mathrm{resid}\left(\mathbf{y},\mathbf{\Phi}_{T^{\ell}}\right).$
\item If $\left\Vert \mathbf{y}_{r}^{\ell}\right\Vert _{2}>\left\Vert \mathbf{y}_{r}^{\ell-1}\right\Vert _{2}$,
let $T^{\ell}=T^{\ell-1}$ and quit the iteration.
\end{enumerate}
\textbf{Output}:

\begin{enumerate}
\item The estimated signal $\hat{\mathbf{x}}$, satisfying $\hat{\mathbf{x}}_{\left\{ 1,\cdots,N\right\} -T^{\ell}}=\mathbf{0}$
and $\hat{\mathbf{x}}_{T^{\ell}}=\mathbf{\Phi}_{T^{\ell}}^{\dagger}\mathbf{y}$.
\end{enumerate}
\end{table}
OMP techniques were recently extended in a manner that allows for adaptively adding or removing sets of column candidates from the estimated list of columns. 
One algorithm that uses this idea is SP algorithm (for more details regarding the SP algorithm, the interested reader is referred to \cite{Wei-Olgica-Subspace-Pursuit}). 
Here, $\mathbf{y}_{r}^{\ell}=\mathrm{resid}\left(\mathbf{y},\mathbf{\Phi}_{T^{\ell}}\right)$ denotes the residual vector of the projection of vector \textbf{y} onto the subspace spanned by $\mathbf{\Phi}_{T^{\ell}}$. 
For completeness, the flow-charts of the OMP and SP algorithms are given in Table 1 and Table 2, respectively. 
Note that the columns of $\mathbf{\Phi}$ are denoted by $\boldsymbol{\varphi}_1,\ldots,\boldsymbol{\varphi}_N$.

The computational complexity of OMP strategies
depends on the number of iterations needed for exact reconstruction:
standard OMP always runs through $K$ iterations, and therefore its
reconstruction complexity is roughly $O\left(KmN\right)$ operations. The same is true of the
SP algorithm, except that for some classes of sensing vectors the complexity can be brought
down to $O\left(mN\, \log \,K\right)$. 

An important open question in CS theory is how to devise structured low-dimensional sensing matrices that can be decoded
with vey low complexity algorithms that exploit the structural properties of the matrices.
Restricting the choices for the sensing matrix to a special class of matrices necessarily introduces a performance loss, and one would like to investigate
the trade-off between the performance loss and complexity of such reconstruction
algorithms. Our results pertaining to these questions are presented in the following sections. 

\section{\label{sec:LDPC_CS}Compressive Sensing Using LDPC Codes}

A binary linear block code with parameters $[m,s,d_{\text{min}}]$, $\mathcal{C}$, is a
$s$-dimensional subspace of an $m$-dimensional vector space over the finite field $\mathbb{F}_2$. Less formally, a code is a collection
of codewords of length $m$ that encode $s$ information bits using $m-s$
parity-check bits.  The parameter $d_{\text{min}}$ is the minimum distance of the code,
defined as the smallest Hamming distance (total coordinate distance) between any pair of
distinct codewords. The code rate is defined as $R=s/m$~\footnote{Note that our notation does not follow the standards in coding literature when it comes to denoting the codelength
and dimension. We use this notation to prevent confusion with the standard notation used in compressive sensing literature.}.

 A set of $s$ basis-vectors of the subspace, arranged
 row-wise, forms a generator matrix of the code, denoted by \textbf{G}.
 A set of $m-s$ basis-vectors of the null-space of $\mathcal{C}$,
 arranged row-wise, forms a parity-check matrix of the code, denoted by 
 \textbf{H}. Clearly, ${\bf c}\in \mathcal{C}$ iff ${\bf H c} = {\bf 0}.$
A low-density parity-check (LDPC) code is a linear block code with at least
one ``sparse'' parity-check matrix \cite{gallager}. The word ``sparse'' is given different
meanings in different contexts. In this work, ``sparse'' refers
to the property that every column or row of parity-check matrix has at most
$w_c$ or $w_r$ non-zero entries, respectively, where $w_c$ and $w_r$ are constants that
do not depend on $m$ or $s$.

\subsection{Sensing Matrix Construction}

Consider a $[m,s,d_{\text{min}}]$ LDPC code $\mathcal{C}$ that does not contain the all-ones codeword. 
We construct a $m \times (N=2^s-1)$ sensing matrix $\mathbf{\Phi}$ in the following manner\footnote{Note that the proposed construction apparently imposes a restriction on the size of the sensing matrix in so far that 
the number of columns $N$ has to be one less than a power of two. But this restriction is not a strict one: it is possible to leave out as many codewords/columns of the matrix as needed to bring it to the size
required by the application at hand. None of the performance guarantees are affected by this modification.}.

First, we convert all non-zero codewords of $\mathcal{C}$ into their \emph{Binary Phase Shift Keying} (BPSK) images, defined via the mapping $+1 \to -1$ and $0 \to +1$.
Subsequently, we normalize each image by $1/\sqrt{m}$. These normalized codewords are used as columns of the matrix $\mathbf{\Phi}$. 
The columns of $\mathbf{\Phi}$ are, as before, denoted by $\boldsymbol{\varphi}_1,\ldots,\boldsymbol{\varphi}_N$, while $\mathbf{c}_1,\ldots,\mathbf{c}_N$ are used to denote their
corresponding codewords. Since LDPC codes are linear codes, one can arrange the columns of $\mathbf{\Phi}$ lexicographically, so that locating one column of $\mathbf{\Phi}$ requires the order of $O(\log N)$
operations.

The crux behind choosing a sensing matrix of this form is as follows: one of the most expensive steps in OMP and SP reconstruction is \emph{correlation maximization}. 
Choosing $\mathbf{\Phi}$ to be composed of codewords of a binary linear code allows one to recast the problem of correlation maximization as the problem of finding the maximum likelihood (ML) codeword.
Furthermore, due to the fact that LDPC codes are linear, each row of $\mathbf{\Phi}$ has $N/2$ negative and $N/2-1$ positive entries. To ensure that the matrix $\mathbf{\Phi}$ ``globally'' mimics the properties of 
a random Bernoulli matrix, one also needs to ensure that the columns of $\mathbf{\Phi}$ (codewords of the code) have weight close to $m/2$. Henceforth, we refer to matrices that have such properties as
"Bernoulli-like" matrices. One has to ensure that these properties hold locally as well, which can be achieved by enforcing similar constraints on subcodes of the code. 

With respect to the first problem, we show that LDPC code with sharply concentrated distance spectra around $m/2$ exist. Regarding the second issue, we demonstrate 
that support weight enumerators \cite{Milenkovic03} can be used to characterize the ``local Bernoulli structure'' of the matrix $\mathbf{\Phi}$.

As before, let $\mathbf{x}\in\mathbb{R}^{N}$ be a $K$-sparse signal, and let $\mathbf{y}\in\mathbb{R}^{m}$
be the CS observation vector.
The maximum correlation between distinct columns of the matrix is denoted by
\[\mu\triangleq\max_{i\neq j}|\langle\boldsymbol\varphi_i,\;\boldsymbol{\varphi}_j\rangle|,\]
where
\[\langle\boldsymbol\varphi_i,\;\boldsymbol{\varphi}_j\rangle=\sum_{l=1}^m{\varphi_{i,l}\varphi_{j,l}}.
\]

The parameter $\mu$ is called the coherence parameter of the matrix $\mathbf{\Phi}$. There exists a fundamental lower bound on the value of the coherence parameter,
\[ \mu \geq \frac{c}{\sqrt{m}}, \]
where $c$ denotes a constant (see~\cite{temlyakov2011} and references therein).

Most performance guarantees of OMP algorithms are expressed in terms of this parameter. The standard OMP algorithm guarantees exact recovery of the vector $\mathbf{x}$ as long as $\mu\leq 1/2K$
\cite{Tropp_IT2004_Greed_is_good}. 
Such a bound, as we show later, holds for sensing ``Bernoulli-like" matrices based on LDPC codes.

It is straightforward to express the coherence parameter of the code-based matrix $\mathbf{\Phi}$
in terms of the Hamming distance $d_H$ between codewords, as shown below:
\begin{eqnarray*}
  \langle\boldsymbol\varphi_i,\;\boldsymbol{\varphi}_j\rangle &=& \sum_{l=1}^m{\varphi_{i,l}\;\varphi_{j,l}} 
  = \sum_{\substack{1\leq l\leq m\\\varphi_{i,l}=\varphi_{j,l}}}{\frac{1}{m}}+\sum_{\substack{1\leq l\leq m
  \varphi_{i,l}  \neq  \varphi_{j,l}}}^m{(-\frac{1}{m})} 
  = 1-\frac{d_{\textrm{H}}(c_i,c_j)}{m}-\frac{d_{\textrm{H}}(c_i,c_j)}{m} \\
  &=& 1-\frac{2d_{\textrm{H}}(c_i,c_j)}{m}.
\end{eqnarray*}
Here, $d_\textrm{H}(c_i,c_j)$ denotes the Hamming distance between the codewords $c_i$ and $c_j$. Note that
if for all $i\neq j$,
\[
\frac{d_{H}(c_i,c_j)}{m}>\frac{1}{2}-\frac{1}{4K},\;\; \mbox{then}\;\; \frac{m-2d_{H}(c_i,c_j)}{m}<\frac{1}{2K},\]
 and if for all $i\neq j$,
 \[
\frac{d_{H}(c_i,c_j)}{m}<\frac{1}{2}+\frac{1}{4K},\;\; \mbox{then,}\;\; \frac{m-2d_{H}(c_i,c_j)}{m}>-\frac{1}{2K}.\]
Consequently, it holds that
\[
\frac{d_{H}(c_i,c_j)}{m}\in\left(\frac{1}{2}-\frac{1}{4K},\frac{1}{2}+\frac{1}{4K}\right)\Rightarrow\left|\left\langle \bm{\varphi}_{i},\bm{\varphi}_{j}\right\rangle \right|<\frac{1}{2K}.\]

Hence, to guarantee exact recovery with the OMP algorithm for all $K$-sparse signals, we need to identify LDPC codes with
\[\frac{1}{2}-\frac{1}{4K}<\frac{d_{H}(c_i,c_j)}{m}<\frac{1}{2}+\frac{1}{4K}, \forall i\neq j.
\]

That such LDPC codes indeed exist is shown in the proposition below.
In the proof, we use the following ensemble of codes from~\cite{Litsyn_Shevelev_LDPC_Distance_Distribution}.

\textbf{Ensemble E}: The parity-check matrix \textbf{H} of the code is chosen with uniform probability from the ensemble of
$(m-s)\times m$ (0,1)-matrices with row sums equal to $w_r$. Such codes are referred to as row-regular codes.

\begin{prop}
\label{prop:LDPC_min_distance}
Consider an LDPC code with codewords $c_{\ell}$, $\ell=1,\ldots,N$, from the Ensemble E with odd $w_r\geq 3$. Let $K\in \mathbb{Z}^{+}$ and let $K,m,N=2^s$ go to infinity with
$m \geq c\,K^2\log N$ for some constant $c$. Then, with probability at least $1-e^{-m\nu}$, for some constant $\nu>0$, one has
\begin{equation}
\frac{1}{2}-\frac{1}{4K}<\frac{d_{H}(c_i,c_j)}{m}<\frac{1}{2}+\frac{1}{4K},\; \forall i\neq j.
\label{eq:relative_distance}
\end{equation}

\end{prop}
\vspace{0.1in}
\begin{remrk}
For even values of $w_r$, since both all-zero and all-one codewords are in the code, (\ref{eq:relative_distance}) is obviously not satisfied. 
In this case, one needs to expurgate all the codewords starting with the symbol $0$ (or, alternatively, starting with the symbol $1$). This does not, 
however, change the asymptotic formula for the rate of the code. Also, note that in this case, the number of rows of the sensing matrix is $K$-times
larger than the smallest number needed for LP reconstruction methods. 
\end{remrk}

To prove Proposition \ref{prop:LDPC_min_distance}, we introduce the following terminology.

Let $\mathcal{C}_m$ be an ensemble of codes of length $m$ defined by parity-check matrices of size $(m-s)\times m$.
For a code $\mathcal{C}\in\mathcal{C}_m$, let
\[B_i(\mathcal{C})=|\{\mathbf{c}\in \mathcal{C}: \text{wt}(\mathbf{c})=i\}|, \;\;i=0,1,\ldots,m,
\]
where wt$(\cdot)$ denotes the Hamming weight. The average ensemble distance distribution is
\[\bar{B}(\mathcal{C}_m)=(\bar{B}_0(\mathcal{C}_m),\bar{B}_1(\mathcal{C}_m),\ldots,\bar{B}_m(\mathcal{C}_m)),
\]
where
\[\bar{B}_i(\mathcal{C}_m):=\bar{B}_i=\frac{1}{|\mathcal{C}_m|}\sum_{\mathcal{C}\in\mathcal{C}_m}{B_i(\mathcal{C})}.
\]

We use the following theorem from \cite{Litsyn_Shevelev_LDPC_Distance_Distribution}.
\begin{thm}
\label{thm:Litsyn_Shevelev_LDPC_min_distance}
Let $\alpha:=(m-s)/s=1-R$. For $\theta\in (0,1)$, the elements of the average distance distribution are of the form
\[b_{\theta}:=\lim_{m\rightarrow \infty} \frac{1}{m}\ln \bar{B}_{\theta m}=H(\theta) + p^{\alpha}_{\theta},
\]
where $H(\theta)$ denotes Shannon's entropy function. For ensemble E, one has the following formula
\[p^{\alpha}_{\theta}=\alpha \ln \left(\frac{1+(1-2\theta)^{w_r}}{2}\right).
\]
\end{thm}
\textbf{Proof of Proposition 1:}
Following the approach of~\cite{Litsyn_Shevelev_LDPC_Distance_Distribution}, we need to prove that with high probability, $b_{\theta}<0$ for any
$\theta\not\in\left(\frac{1}{2}-\frac{1}{4K},\frac{1}{2}+\frac{1}{4K}\right)$. If $b_{\theta}<0$, the average number of codewords of weight $\theta\,m$ in the
ensemble goes to zero as the length of the codes increases.

Let $\theta^{-}=\frac{1}{2}-\frac{1}{4K}$ and
choose $\epsilon>0 $ such that $\theta=\frac{1-\epsilon}{2}$. Then
\begin{eqnarray}
\label{eq:entropy_theta}
  H\left(\theta\right) &=& \log 2-\frac{1-\epsilon}{2}\log\left(1-\epsilon\right)-\frac{1+\epsilon}{2}\log\left(1+\epsilon\right)\nonumber
   = \log2-\frac{1-\epsilon}{2}\left(-\epsilon+\frac{\epsilon^{2}}{2}+o\left(\epsilon^{2}\right)\right)\nonumber\\
   && -\frac{1+\epsilon}{2}\left(\epsilon+\frac{\epsilon^{2}}{2}+o\left(\epsilon^{2}\right)\right)\nonumber
   =  \log2-\frac{3\,\epsilon^{2}}{2}+o\left(\epsilon^{2}\right).
\end{eqnarray}

With $\theta^{-}=\frac{1}{2}-\frac{1}{4K}$ or $\epsilon^{-}=\frac{1}{2K}$, (\ref{eq:entropy_theta}) becomes $H\left(\theta^{-}\right)=\log2-\frac{3}{8K^{2}}+o\left(\frac{1}{K^{2}}\right).$
Note that $H\left(\theta\right)\leq H\left(\theta^{-}\right), \forall \theta<\theta^{-}$.
Then, for $\alpha=1-R$,
\begin{eqnarray*}
  b_{\theta} &\leq& b_{\theta^{-}}=H\left(\theta^{-}\right)+p_{\theta^{-}}^{\alpha}
   = \log2-\frac{3}{8K^{2}}+o\left(\frac{1}{K^{2}}\right)-\left(1-R\right)\log2\\
   && +\left(1-R\right)\log\left(1+\left(\frac{1}{2K}\right)^{w_{r}}\right)
   = R\log2-\frac{3}{8K^{2}}+o\left(\frac{1}{K^{2}}\right)+o\left(\frac{1}{K^{w_{r}}}\right).
\end{eqnarray*}

Similarly, one can prove that $\forall \theta > \theta^{+}=\frac{1}{2}+\frac{1}{4K}$,
\[b_{\theta}\leq b_{\theta^{+}}
= R\log 2 -\frac{3}{8K^2}+o\left(\frac{1}{K^{2}}\right)+o\left(\frac{1}{K^{w_{r}}}\right).
\]

Therefore, if $
R<\frac{1}{K^{2}}\frac{1}{8\log2},$
or equivalently, if
$m>(8/3)\log\left(2\right)K^{2}\log N,$
then $\forall \theta\not\in \left(\frac{1}{2}-\frac{1}{4K},\frac{1}{2}+\frac{1}{4K}\right)$, $b_{\theta}<0$, when $K,m,N$ go to infinity. 
This proves that the average ensemble's relative distance lies within the interval $\left(\frac{1}{2}-\frac{1}{4K},\frac{1}{2}+\frac{1}{4K}\right)$ 
with probability at least $1-e^{-m \nu}$, for some constant $\nu>0$. Therefore, according to the probabilistic method, there exists at least one code 
which exceeds the average, and this proves the claimed results.

\begin{remrk}
Proposition 1 suggests that when $m \geq c \, K^{2}\log N$, our
construction ensures that the maximum correlation between distinct
columns is small, i.e., $\mu\le1/2K$. In fact, the following proposition
about random Bernoulli sensing matrix shows that our LDPC based construction
is asymptotically as good as the random Bernoulli counterpart.\end{remrk}

\begin{prop}
\label{prop:perf-random-Bernoulli}Let $\bm{X}\in\mathbb{R}^{m\times N}$
whose entries are randomly drawn from the Bernoulli distribution with
parameter $p=1/2$. Let $\bm{x}_{i}$ be the $i^{th}$ column of the
matrix $\bm{X}$. Let $K,m,N$ approach infinity simultaneously. Then
if and only if $m\ge cK^{2}\log N$ for some constant $c>0$, one
has $\underset{i\ne j}{\max}\frac{1}{m}\left|\left\langle \bm{x}_{i},\bm{x}_{j}\right\rangle \right|\le\frac{1}{2K}$
with probability close to one.\end{prop}
\textbf{Proof of Proposition 2}
See Appendix \ref{sub:Proof-of-random-Bernoulli}.

We can now bound the RIP parameter of the code-based matrix $\mathbf{\Phi}$. Applying the
Gershgorin circle theorem \cite{Book_Gershgorin_circle} to the matrix $\mathbf{A}=\mathbf{\Phi}_{I}^{*}\mathbf{\Phi}_{I}$ shows that all
the eigenvalues of \textbf{A} lie in a disc $D\left(1,r\right)$ centered at one and with radius $r$, where
\begin{eqnarray*}
r&=&\max_{i}\sum_{j\neq i}{|\langle\boldsymbol{\varphi}_i,\boldsymbol{\varphi}_j\rangle|} ,\\
&<& K\mu.
\end{eqnarray*}
Therefore, every eigenvalue $\lambda$ of \textbf{A} satisfies
\[1-K\mu<\lambda <1+K\mu.
\]
Hence, if $\mu\leq 1/2K$, it is easy to see that the RIP parameter satisfies
\[\delta_K\leq 1/2.
\]

Alternatively, one can also easily show a more general result that any matrix with coherence parameter $\mu$ and sparsity parameter $K$
satisfies the RIP with constant $\delta=(K-1)\mu$. For the matrices based on LDPC codes, this result also implies $\delta<1/2$

\subsection{Higher Hamming Weights and the RIP}

Henceforth, we consider two classes of sensing vectors $\mathbf{x}$. The first class, referred to as \emph{binary} sensing vectors, has the property
that all non-zero entries of the vector are equal to one. The second class has the property that the non-zero entries are drawn independently, following
a standard Gaussian distribution. Such vectors will be called \emph{Gaussian} sensing vectors.
 
We describe next an interesting connection between code invariants, known as higher weights and support weight distributions, 
and the RIP property restricted to \emph{binary vectors} \textbf{x} only. Furthermore, this example illustrates that local Bernoulli-like matrix
properties depend on the support weights of the LDPC code used to construct $\mathbf{\Phi}$.

Consider four codewords of a linear code of length $m=6$ and their BPSK images, shown below
\[
0\;0\;0\;0\;0\;0,\;\; 1\;1\;1\;1\;0\;0, \;\; 0\;0\;1\;1\;0\;0, \;\; 1\;1\;0\;0\;0\;0,
\]
\begin{equation}
\begin{split}
&+1\;+1\;+1\;+1\;+1\;+1, \;-1\;-1\;-1\;-1\;+1\;+1, \\
&+1\;+1\;-1\;-1\;+1\;+1, \; -1\;-1\;+1\;+1\;+1\;+1. \notag
\end{split}
\end{equation}
Together, these four codewords form a linear code that may represent a \emph{subcode} of a
larger code. The  effective length of the subcode is four, while the length of the code is six -- the last two coordinates are fixed to zero. 

Since a subcode of a linear block code that contains $2^s$ codewords, with $s \geq 2$, has
the property that the total number of zeros and ones for each coordinate in the effective support
is $2^{s-1}$, one can partition the set of codewords into two subsets with $2^{s-2}$ zeros and ones per
effective coordinate. Consequently, summing up the BPSK images of the words in these two partitions
will produce identical vectors. Arbitrary translations of those vectors cannot be distinguished by any compressive sensing reconstruction algorithm.
In the example above, the sum of the first two codewords equals the sum of the last two vectors.

The submatrix of $\mathbf{\Phi}$ induced by codewords corresponding to a subcode contains sub-blocks of identical
entries (equal to $+1$) at positions outside of the effective support of the subcode. As a consequence, if there exist many subcodes of small effective
support size and relatively large dimension, the matrix $\mathbf{\Phi}$ will not be a Bernoulli-like matrix in a local sense. In the example above,
two out of six coordinates have a fixed value equal to one - and consequently, there exists a $2 \times 2$ submatrix of all ones in the induced sensing matrix.

Consequently, the ``quality'' of a code for binary vector CS can be partly characterized in terms of the support weight distribution: there should be 
as few as possible subcodes in the code of small dimension and small effective support weight. Note that in addition to subcodes, there may exist other ``indistinguishable'' collections of 
codewords, but these are in general extremely hard to characterize for a given LDPC code.

None of the problems described above are encountered for the case of real-valued sensing vectors, since the probability of having the same weighting coefficients for two different subsets of columns is
negligible.

There exists a large body of work regarding the \emph{higher weight and support weight distribution}  of codes (see \cite{Milenkovic03} and references within), which provide
information on the smallest support of a subcode of given dimension and the number of such subcodes. In the lemma that follows, we use $d_r$ to denote 
the $r$-th generalized Hamming weight of $C$, i.e., the size of the smallest support of an $r$-dimensional subcode of $C$. Additionally, $H(\gamma)$ denotes Shannon's binary
entropy function with parameter $0<\gamma<1$.

\begin{lemma}\cite{Wei91}
Let the rate of a family of codes satisfy  $R<1-2^{-r}$ and let $\min\{\gamma \log_2(2^r-1);r(\gamma+(r-1)/n)\}<r(1-R)-H(\gamma)$, for some
non-negative integer parameter $r$. Assume that
the codelength $m$ of the codes is sufficiently large. Then almost all $[m,mR]$ codes $C$ have $d_r(C) \geq \gamma m$.
\label{lem:Wei_91}
\end{lemma}

The following proposition provides a necessary condition on the rate of an LDPC code to satisfy the \emph{binary vector} RIP property with 
$\delta_{2K}<\sqrt{2}-1$, which guarantees exact recovery of $K$-sparse signals under LP reconstruction \cite{Candes2008_Paris_RIP_CS}. 

\begin{prop}
\label{prop:Necc-Condition-Rate}
Consider an LDPC code-based sensing matrix $\mathbf{\Phi}$ and a sparse binary input signal. The necessary condition on the rate of the corresponding 
LDPC code for $\mathbf{\Phi}$ to satisfy the binary vector RIP property with parameter $\delta_{K}<\sqrt{2}-1$ is
\[R<1-(1-\frac{\sqrt{2}}{K})\frac{\log_2(K-1)}{\log_2(K)}-\frac{H(\sqrt{2}/K)}{K}.
\]
\end{prop}
\textbf{Proof of Proposition 3:}
We consider a binary vector \textbf{x} whose $K=2^r$ non-zero entries are at positions indexing $K$ columns of $\mathbf{\Phi}$ that form a translation of an $r$-dimensional subcode of $C$. 
Since for a subcode with effective length $d_r$, $m-d_r$ coordinates are equal to zero, it follows that
\[\|\mathbf{\Phi x}\|^2_2=(m-d_r)K^2\leq (1-\gamma)\, m\, K^2.
\]
Then, for $\mathbf{\Phi}$ to satisfy the binary RIP property with parameter $\delta_K$, one needs
\[(1-\gamma)K^2<(1+\delta _{K})K,\]
or
\[\gamma > 1-\frac{1+\delta_{K}}{K}.
\]
Thus, we require
\begin{equation}
\gamma>1-\frac{\sqrt{2}}{K}.
\label{eq:delta-deltaK-K}
\end{equation}
Invoking the conditions of the Lemma \ref{lem:Wei_91}, one can enforce
\[\gamma \log_2(K-1)<\log_2(K)(1-R)-H(\gamma),
\]
or equivalently,
\begin{equation}
R<1-\gamma \frac{\log_2(K-1)}{\log_2(K)}-\frac{H(\gamma)}{\log_2(K)}.
\label{eq:rate-delta-K}
\end{equation}
From (\ref{eq:delta-deltaK-K}) and (\ref{eq:rate-delta-K}), one can obtain necessary conditions on the LDPC code rate $R$ that ensure that the sensing 
matrix satisfies the binary RIP property with parameter $\delta_K<\sqrt{2}-1$ as stated in Proposition \ref{prop:Necc-Condition-Rate}.
From the previous result, it can be seen that the code rate needs to be very small in order to ensure that a matrix with sufficiently strong RIP constant exists. 
However, this requirement is only a necessary condition for the binary RIP property, while the RIP property is a sufficient condition for exact CS reconstruction.
In fact, the code-based sensing matrix still performs very well for much higher code rates, as shown in the Simulation Results section.

\section{\label{sec: Algorithm} Algorithms}

One of the main steps of greedy reconstruction algorithms is to identifying one (or multiple) columns of $\mathbf{\Phi}$ that exhibit maximum correlation with the vector \textbf{y}. 
Usually, such a step is implemented via exhaustive search, which meant that at each iteration of the procedure, one needs to compute $N$ inner products and order them. 
The idea behind the CS technique based on LDPC sensing matrices is that columns of maximum correlation correspond to the most likely codewords, and can hence be efficiently identified using iterative decoding methods such as BP~\cite{gallager}. It is important to point out that BP decoders in this context operate in a vastly different regime than standard decoders: 
they must be able to handle high \emph{interference noise} and potentially identify \emph{lists} of most likely codewords for use in subspace pursuit techniques. Hence, BP methods have to be generalized
to accommodate for signal interference through biasing and reinforcement. 

The main contributions of our work in the context of reconstruction algorithm design are twofold. First, we propose a \emph{novel} list decoding BP approach for binary sensing vectors. The idea behind the list decoder
is to "bias" the components of the measurement vector $\mathbf{y}$ in order to identify the most likely columns of the sensing matrix to have contributed to the observation. This biasing is coupled with a model for
codeword interference borrowed from the theory of multiuser communication systems~\cite{muller}. Second, we present the first modifications of $M$ best codewords decoding, multiple-basis BP and reinforced BP in the context of 
compressive sensing reconstruction. All these algorithms are of independent interest in iterative decoding theory for interference channels. 

In what follows, we start with a brief introduction of the BP algorithm and the reinforced BP algorithm. We then proceed to describe the list decoding BP method for binary sensing vectors. 
We conclude the section by describing how to adapt three generalizations of the BP algorithm to real-valued CS reconstruction.

\subsection{The BP Algorithm}
\label{sec:bp}
Decoding of LDPC codes is standardly performed in terms of iterative algorithms that exchange reliability messages between the variable nodes and the check
nodes of a bipartite Tanner graph representation of the code. Such algorithms are collectively known as a message passing (MP) algorithms, and they include
the BP algorithm, and in particular, the
sum-product algorithm. 
For completeness, we describe the sum-product algorithm of~\cite{mackayetal96} below. 

Throughout this section, we use $\mathcal{G}(\mathbf{H})$ to denote a suitably chosen bipartite Tanner graph of 
an LDPC code $\mathcal{C}$.
The columns of \textbf{H} are indexed by the elements of the set
of variable nodes (left hand side nodes), $\mathcal{V_L}$, of
$\mathcal{G}(\mathbf{H})$, while the rows of \textbf{H} are indexed by the set of
check nodes (right hand side nodes) $\mathcal{V_R}$ of $\mathcal{G}(\mathbf{H})$.

Let us denote the set of variable nodes neighboring a check node
indexed by $a$ by $\mathcal{N}(a) \equiv \left\{ b: H_{a \, b}=1
\right\}$. Similarly, denote the set of check nodes neighboring a
variable node indexed by $b$ by $\mathcal{M}(b) \equiv \left\{ a:
H_{a \, b}=1 \right\}$; the notation $\mathcal{M}(b) \backslash a$ is
reserved for the set $\mathcal{M}(b)$ excluding the check node
$a$.

 During BP decoding, two types of probability messages
$q_{ab,x_b}$ and $r_{ab,x_b}$ are exchanged between a check node
$a$ and a variable node $b$ if and only if the parity-check matrix
has a nonzero entry at the respective position, i.e., if and only if $H_{a \, b}=1$. 
Here, the
message $q_{ab,x_b}$ represents the probability that variable $b$
of the transmitted codeword is $x_b$ ($x_b \in \{{0,1\}}$) given
the information provided by all check nodes incident to variable
node $b$, excluding $a$ and the channel output corresponding to
the variable $x_b$. The message $r_{ab,x_b}$ represents the
probability of check node $a$ being satisfied, provided that the
variable $b$ has value $x_b$ while all other variables incident to
the check node have a separable distribution. The probabilities
$q_{ab,x_b}^{(l)}$, where the superscript $(l)$ indicates the
iteration index, are initialized to the prior channel
probabilities, according to $q_{ab,0}^{(0)}=p_{b,0}=P(x_b=0)$, and
$q_{ab,1}^{(0)}=p_{b,1}=P(x_b=1)=1-p_{b,0}$.

The \emph{horizontal step} of the $l$-th iteration of the
algorithm is initiated by computing the probabilities
$r_{ab,\,x_b}^{(l)}$. This is accomplished in terms of the \emph{fast implementation}
approach described in~\cite{mackayetal96}: first, $\delta r_{ab} = r_{ab,\, 0}^{(l)} -
r_{ab,\, 1}^{(l)}$ is computed according to
\begin{eqnarray}
\delta r_{ab} = \Big( \prod_{b' \in \mathcal{N}(a)} (q_{ab',\,
0}^{(l-1)} - q_{ab',\,1}^{(l-1)}) \Big) \ / \ (q_{ab,\, 0}^{(l-1)}
- q_{ab,\,1}^{(l-1)}).
\label{eq:horizontal_step_convenient_implementation}
\end{eqnarray}

The horizontal step is terminated by computing $r_{ab,\, 0} =
\frac{1}{2} (1+\delta r_{ab})$ and $r_{ab,\, 1} = \frac{1}{2}
(1-\delta r_{ab})$. 

During the \emph{vertical step} of the $l$-th iteration of the BP
algorithm, the values of $q_{ab,\,0}$ and $q_{ab,\,1}$ are updated
as follows. First, the \emph{pseudoposterior probabilities} of
$x_b$ being zero or one are found according to
\begin{eqnarray}
& q_{b,\,0}^{(l)} &= \alpha_{b} p_{b,\,0} \prod_{b \in \mathcal{M}
(b)} r_{ab,\,0}^{(l)}\nonumber\\ & q_{b,1}^{(l)} &= \alpha_{b}
p_{b,1} \prod_{b \in \mathcal{M} (b)} r_{ab,1}^{(l)},
\label{eq:pseudoposterior}
\end{eqnarray}
where $\alpha_{b}$ is chosen so as to ensure
$q_{b,\,0}^{(l)}+q_{b,\,1}^{(l)}=1$. The probabilities described
above are used for tentative decoding. The algorithm stops if the
hard-decision values of the pseudoposterior probabilities
represent a valid codeword of the LDPC code, i.e., a word $\mathbf{c}$ such that
$\mathbf{H}\mathbf{c}^T=\mathbf{0}$.  

Otherwise, from the pseudoposterior
probabilities, the values of $q_{ab,\,0}$ and $q_{ab,\,1}$ are
calculated according to the equations:
\begin{eqnarray}
 q_{ab,\,0}^{(l)} &=& \alpha_{ab} p_{b,\,0} \quad q_{b,\,0}^{(l)} / r_{ab,\,0}^{(l)}\nonumber\\
 q_{ab,\,1}^{(l)} &=& \alpha_{ab} p_{b,\,1} \quad q_{b,\,1}^{(l)} / r_{ab,\,1}^{(l)},
\label{eq:vertical_step}
\end{eqnarray}
where $\alpha_{ab}$ is chosen so that $q_{ab,\,0}^{(l)}+
q_{ab,\,1}^{(l)}=1$. These probabilities are passed on to the
horizontal step of the next iteration.

\subsection{Reinforced BP}

The BP algorithm is known to efficiently reach a near-optimal solution whenever the variables are ``biased'' or directed towards such a solution. 
The idea behind reinforced BP is to provide an additional bias towards the correct solution by combining properly weighted messages from the past and 
the present~\cite{Braunstein_ReinforcedBP}. The Tanner graph of the code and the reinforced decoder contains an additional level of nodes, the
reinforcement nodes, that serve to store messages passed on by the algorithm in the past. We restrict our attention to memory-one reinforcement strategies,
were only messages from one previous iteration are stored. The reinforced BP algorithm is described in detail below.

RBP modifies the messages from variable nodes to check nodes: the idea is to introduce an extra term into (\ref{eq:vertical_step}). This is accomplished as 
follows. Define the marginal function of variable $b$ at iteration $l$ as
\[
g^{(l+1)}(b)\propto\prod_{a\in\mathcal{M}(b)}r_{ab,x_b}^{(l)}.
\]
In RBP, the horizontal step (\ref{eq:horizontal_step_convenient_implementation}) does not change compared to standard BP, 
whereas the vertical step (\ref{eq:vertical_step}) is modified as described below:
\begin{equation}
q_{ab,\,x_b}^{(l)} = \alpha_{ab} (g^{(l)}(b))^{\psi(l)}p_{b,\,x_b} \quad q_{b,\,x_b}^{(l)} / r_{ab,\,x_b}^{(l)}.
\end{equation}
Here, $x_b\in\{0,1\}$, $\alpha_{ab}$ is a normalization factor and $\psi(l):[0,1]\rightarrow [0,1]$ is a non-decreasing function, often chosen to be of the form
\[\psi(l)=1-\psi_0\psi_1^l,
\]
where $\psi_0,\psi_1$ are in $[0,1]$. The speed of convergence of reinforced BP depends on the values of $\psi(l)$, and therefore, these parameter have to be carefully chosen. 

\subsection{\label{sec:ListDecodeBP}List Decoding BP, BP-OMP, and BP-SP for Binary Vectors}

We focus our attention on a new class of BP decoders for binary vectors $\mathbf{x}$ only. 
The BP decoder has the purpose to identify the column of $\mathbf{\Phi}$ that maximizes the correlation with the measurement vector $\mathbf{y}$. 
In this setting, two observations are at place. 

First, when $K=2$, both columns (codewords) of $\mathbf{\Phi}$ used in the superposition have the same correlation with \textbf{y}. This is a particularly difficult setup for BP decoding, since the measurement vector 
$\mathbf{y}$ may fall in the middle of the decision regions for the two codewords, but not converge to any of these two codewords. This motivates analyzing strategies for \emph{biasing} the estimates for 
certain coordinates in $\mathbf{y}$ so as to move the vector away from the boundary region. Biasing may be performed using reinforced BP, but as we show shortly, there is a much simpler and more 
efficient way to perform this biasing directly when the sensing vector is binary.

Second, whenever one tries to identify the column with largest correlation, all the remaining columns in the superposition act as interference. For large $K$, this interference may be very severe. 
Hence, the question of interest is to find an adequate model for the interference and investigate the performance of BP decoders for a very non-standard operational regime - namely, a regime significantly 
below the code's designed threshold.

We start with the description of the biasing scheme for binary vectors. For illustrative purposes, we first describe the scheme for the case $K=2$. We observe that for $K=2$, the entries of the measurement vector \textbf{y} must lie in the set $\{-2,0,2\}$. Whenever $y_i=+2$ or -2, both entries of the two columns of $\mathbf{\Phi}$ in the superposition must be $+1$, or $-1$, respectively. For zero entries of \textbf{y}, one column must take the value $-1$ and the other must take the value $+1$ at that given coordinate. Therefore, one simple idea is to bias the $\pm 2$ entries toward some large value, ideally $\pm \infty$, to ensure that the BP algorithm decodes the corresponding bits correctly. One can further bias one of the remaining zero entries of \textbf{y} either towards a large positive or negative value in order for the BP algorithm to decode it to $+1$ (or $-1$), respectively.

For the more general case, whenever $y_i= +K$ or $-K$, one can bias the corresponding entries towards some large positive or negative value, respectively. When $y_i\neq\pm K$, at least one of the columns must have the value $+1$ and at least one other column must have the value $-1$ at that coordinate. Note that biasing of values $\neq \pm K$ may be problematic, since either of the two values $+1$ and $-1$ is equally plausible for the
ML codeword. Consequently, one should run the BP algorithm for different choices of biasing coordinates, which both increases the complexity of the scheme and makes the search list larger.
Consequently, we restrict our attention to biasing $\pm K$ and $\pm (K-1)$ values only, and limit the later biasing to only a few randomly chosen entries satisfying the required constraint. This leads only 
to a constant increase in complexity of the scheme. The generated list of potential columns (codewords) has to be expurgated from vectors that correspond to words obtained by 
divergent BP runs and repeated vectors. If the output list has more than $K$ codewords, one outputs only $K$ codewords from the list that have largest correlations with $\mathbf{y}$. Otherwise, 
if the list has less than $K$ entries, one outputs the whole list augmented by a set of randomly chosen columns that make the list of size $K$.

As stated earlier, whenever one wants to identify a column $\boldsymbol{\varphi}_k$ in the superposition, the remaining columns act as interference.

The algorithm outlined above is summarized in \emph{Algorithm 3}. 
The list decoding BP algorithm can be combined with standard OMP reconstruction techniques, as summarized in \emph{Algorithm 4}. Furthermore, the SP algorithm can be easily modified to include
the biased list-based BP decoding strategy as its correlation maximization step so as to reduce its overall complexity. 

The performance of the BP list decoder can be assessed in several different ways. One way is to declare reconstruction success if at \emph{least one} of the columns of $\mathbf{\Phi}$ has been identified correctly. The second measure can be the probability of identifying \emph{all} columns of $\mathbf{\Phi}$ within the given superposition.
The larger the size of the list, the larger the probability of finding the correct codewords in the list. However, increasing the list size also means larger computational complexity, since one has to run the BP algorithm for each element in the list.

\begin{algorithm}[H]
\label{alg: ListDecode_description}
\noindent \textbf{Input}: $K$, \textbf{H}, $\mathbf{y}$, biasing list size $L$, large, a large positive biasing value $B$.

\begin{itemize}
\item  Construct the list $\mathbf{RX}=(\mathbf{rx}^{(1)},\mathbf{rx}^{(2)},\ldots,\mathbf{rx}^{(L)})$, where
\begin{itemize}
  \item $\mathbf{rx}^{(1)}=\mathbf{y}$
  \item $\mathbf{rx}^{(2)}$ is the vector $\mathbf{y}$ with entries $\pm K$ reset to $\pm B$
  \item Randomly generate an index set $\mathcal{J}$ of size $(L-2)/2$, where $rx_{j} = \pm (K-1),\forall j\in \mathcal{J}$.
  \item For each $j \in \mathcal{J}$, bias the $j^{th}$ coordinate of the vector $\mathbf{rx}^{(2)}$ to $+B$ and $-B$ to get  $\mathbf{rx}^{(2l+1)}$ and $\mathbf{rx}^{(2l+2)}$, respectively, ($l=1,2,\ldots,(L-1)/2$).
\end{itemize}
\item Run the BP algorithm for $\mathbf{rx}^{(i)}$ and output a binary column vector $\mathbf{\hat{v}}_i$ $(i=1,\ldots,L)$. Delete all the words which do not satisfy $\mathbf{H} \mathbf{\hat{v}}_i=0$ and delete all the repetitions in the list.
\item Output at most K words $\mathbf{\hat{v}}_{i}$ whose BPSK images have largest correlations with $\mathbf{y}$.
\end{itemize}

\noindent \textbf{Output}: A potential list of columns of $\mathbf{\Phi}$ that contributed to $\mathbf{y}$.

\caption{\label{alg:List-Decoding-Algorithm}Biased List-based BP (BLBP Algorithm}
\end{algorithm}

It is straightforward to see that the computational complexity of the BP-OMP algorithm equals $O\left(KLm\right)$. 
This follows from the fact that BP runs in time proportional to the length of the code, provided that the number of iterations is fixed. 
For fixed list size $L$, which is the setup we used, the complexity equals $O\left(Km\right)$. It is worth pointing out that BP is a suboptimal algorithm, and that 
its exact analytical performance characterizations is still not known. 

\begin{algorithm}[H]
\label{alg:BP-OMP-alg-description}
\noindent \textbf{Input}: $K$, \textbf{H}, $\mathbf{y}$, $L$, $B$.

\noindent \textbf{Initialization}: $\mathbf{rx}=\mathbf{y}$

\noindent \textbf{Iteration}: Run the iteration $K$ times
\begin{itemize}
\item Run the Biased list-based BP algorithm and output a list of potential columns
\item Pick the codeword \textbf{v} whose BPSK image has largest correlation with \textbf{rx}
\item Update: $K=K-1$, $\mathbf{rx}=\mathbf{y}-\textrm{BPSK}(\mathbf{v})$
\end{itemize}
\textbf{Output}: The list of $K$ columns of $\mathbf{\Phi}$ in the superposition.

\caption{\label{alg:BP-OMP-Algorithm}BP-OMP Algorithm}

\end{algorithm}

\subsection{List Decoding BP Methods for Gaussian Sensing Vectors}

We describe next how to implement list decoders for real-valued sensing vectors. The biasing methods we described in section \ref{sec: Algorithm} can not be used for vectors $\mathbf{x}$ that are non-binary, due to 
the fact that biasing is contingent on the vector $\mathbf{y}$ having bounded integer values. To mitigate this problem, we propose using three different approaches, previously described in the context of graphical 
model analysis only. All these methods
may be seen as schemes that bias the BP decoder performance in a way that allows one to generate multiple candidates for the ML codeword. The methods are the $M$ Most Probable Configurations (MMPCs) algorithm~\cite{yanover_findingm_2003}, the multiple-basis belief propagation (MBBP) algorithm introduced by one of the authors in~\cite{MBBP_Olgica}, and the previously described reinforced BP algorithm. 

The idea behind the MMPCs algorithm can be summarized as follows. First, the most probable configuration $\mathbf{x}_1$ is found by running the BP algorithm. The symbol of largest marginal probability
is consequently frozen to its complement. In the second step, the algorithm performs a search for the next best set of marginal values with the frozen symbol, and then recalculates the marginal probabilities. Freezing
the symbol value allows one to avoid finding the most probable configuration once again in the second round of decoding. The output of the second step of the decoding process is denoted by  $\mathbf{x}_2$. 
Similarly, $\mathbf{x}_k$ is found in such a way that it differs from $\mathbf{x}_1,\mathbf{x}_2,\ldots,\mathbf{x}_{k-1}$ in at least one coordinate. The interested reader is referred to~\cite{yanover_findingm_2003} for 
more details regarding this algorithm and several illustrative examples. The steps of the method, referred to as the MMPCs algorithm, were adapted from
 \cite{yanover_findingm_2003} and are summarized in Algorithm \ref{alg:BMMF-Algorithm}.

On the other hand, the MBBP algorithm utilizes several parity-check matrices in parallel to identify columns of $\mathbf{\Phi}$ that have largest correlation with $\mathbf{y}$. The different parity-check 
matrices are carefully chosen so as to bias the decoders in as many different directions leading to a number of near-optimal solutions~\cite{MBBP_Olgica}. This can be attributed to the fact that parity-check matrices with 
different degree distributions are employed, which may perform differently for the same interference model.

The steps of the MBBP algorithm adapted for CS are summarized in Algorithm 4.

Note that both the MMPCs and MBBP algorithms may be applied to binary sensing vectors as well. The computational complexity of the algorithms is $O(Mm)$ and $O(Lm)$, where in the latter case, $L$ is used to
denote the number of parallel decoders (in analogy with the biased BP algorithm for binary vectors where $L$ denotes the number of runs of the BP algorithm). Note that MBBP is performed in parallel for all decoders,
and theoretically takes only $O(m)$ operations, but its storage complexity is $L$ times higher due to the need for storing multiple parity-check matrices.

Finally, the MMPCs and MBBP algorithms may be combined with the OMP and SP algorithm for compressive sensing reconstruction. One problem associated with the use of these algorithms is that each run of MMPC may produce less than $K$ vectors since the BP algorithm may not converge. To solve this problem, one may randomly pick columns of $\mathbf{\Phi}$ to fill up the list, as described for binary sensing vectors.

\section{Simulation Results} \label{sec:simulation}

Throughout our simulation section, we follow a common practice in coding theory to illustrate the performance of different decoding/CS reconstruction algorithms on one fixed code. In our simulations, the code is a length $160$ progressive edge growth (PEG) row-regular LDPC code~\cite{hu2003}. The dimension of the code is $s=10$, resulting in a coderate of $1/16$. In order to highlight the drawbacks and the advantages of different algorithms, we 
will describe our findings for other code parameters as well. 
For example, we may change the length of the code, or change its row- or column-degree. 

In all our simulation, we model the interference as a zero-mean Gaussian variable with variance $\sigma_{k}^2=\max(|\mathbf{y}|)(K-1)/K$.

\begin{algorithm}[H]
\noindent \textbf{Input}: $K$, $\mathbf{H},\mathbf{\Phi}$, $\mathbf{y}$.

\textbf{Initialization}
\begin{eqnarray*}
\text{SCORE}_1(i,j)&=&\max_{\mathbf{x}:\mathbf{x}(i)=j}\text{Pr}(X=\mathbf{x}|\mathbf{y}),\\
\mathbf{x}_1(i)&=&\arg\max_j \text{SCORE}_1(i,j),\\
\text{CONSTRAINTS}_1&=&\emptyset,\\
\text{USED}_2&=&\emptyset.
\end{eqnarray*}
\textbf{Iteration}: For $k=2$ to $K$:
\begin{eqnarray*}
\text{SEARCH}_k&=&(i,j,s<k:\mathbf{x}_s(i)\neq j,\\
&& (i,j,s)\not\in \text{USED}_k),\\
(i_k,j_k,s_k)&=&\arg\max_{(i,j,s)\in \text{SEARCH}_k} \text{SCORE}_s(i,j),\\
\text{CONSTRAINTS}_{k}&=&\text{CONSTRAINTS}_{s_k}\cup \{(\mathbf{x}(i_k)=j_k)\},\\
\text{SCORE}_k(i,j)&=&\max_{\mathbf{x}(i)=j,\text{ CONSTRAINTS}_{s_k}}\text{Pr}(X=\mathbf{x}|\mathbf{y}),\\
\mathbf{x}_k(i)&=&\arg\max_j \text{SCORE}_1(i,j),\\
\text{USED}_{k+1}&=&\text{USED}\cup \{(i_k,j_k,s_k)\},\\
\text{CONSTRAINTS}_{s_k}&=&\text{CONSTRAINTS}_{s_k}\cup\{(\mathbf{x}(i_k)\neq j_k)\},\\
\text{SCORE}_{s_k}(i,j)&=&\max_{\mathbf{x}(i)=j,\text{ CONSTRAINTS}_{s_k}}\text{Pr}(X=\mathbf{x}|\mathbf{y}).
\end{eqnarray*}
\noindent \textbf{Output}: $K$ vector with highest correlation with \textbf{y}
\caption{\label{alg:BMMF-Algorithm}The MMPCs Algorithm}
\end{algorithm}

\begin{algorithm}[H]
\noindent \textbf{Input}: $K$, $\mathbf{H}_1,\mathbf{H}_2,\ldots,\mathbf{H}_p$, $\mathbf{y}$

\begin{itemize}
\item For $i=1$ to $p$:
Run the standard BP algorithm on $\mathbf{y}$ using the matrices $\mathbf{H}_i$, $i=1,\ldots, L$, and output a word $\mathbf{v}_i$ for each decoder if it converges
\item Delete all the column vectors which do not satisfy $\mathbf{H} \mathbf{\hat{v}}_i=0$
\item Output the codewords whose BPSK image have the highest correlation with the received vector
\end{itemize}

\noindent \textbf{Output}: One potential column of $\mathbf{\Phi}$ in the superposition.

\caption{\label{alg:MBBP-Algorithm}MBBP Algorithm}

\end{algorithm}

 
We start with an analysis of the performance loss of LDPC code based sensing matrices, as compared to random-like Bernoulli matrices. For our simulations, we generated two types of  parity-check matrices. 
One class of matrices is generated randomly with uniform row-degree four and with columns of degree four or three 
(row-regular matrices) and the other class is generated randomly with uniform column-degree three and with rows of degree three and four (column-regular matrices). One can easily see that 
with row-regular parity-check matrices, the all-one codeword is in the codebook. Therefore, there exist 
codewords $\mathbf{c}_1,\mathbf{c}_2$ in the codebook such that $\mathbf{c}_1+\mathbf{c}_2=1$, or equivalently, codewords whose BPSK images cancel out. 
To avoid this problem, for row-regular parity check matrix, we only use the codewords whose first coordinate is zero in order to form the sensing matrix $\mathbf{\Phi}$. This coordinate 
is subsequently deleted from all the codewords. Hence, in this case, the code originally had dimension $11$, and was subsequently shortened.

\begin{figure}[ht]
\centering
\subfigure[Performance of the OMP algorithm with LDPC code based sensing matrices $\mathbf{\Phi}$.]{
\includegraphics[width=0.45\textwidth]{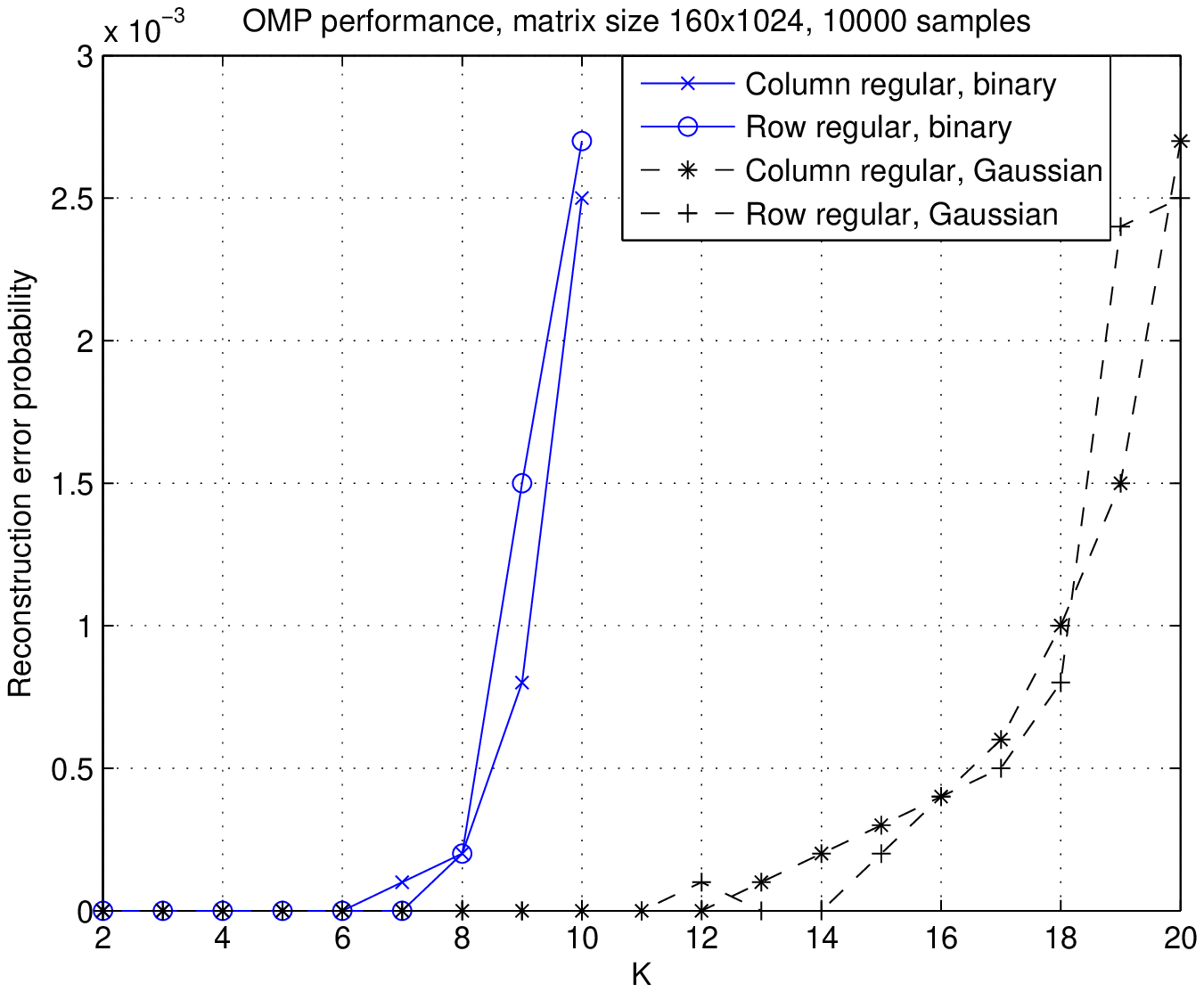}
\label{fig:Standard_OMP}
}
\subfigure[Performance of the SP algorithm with LDPC code based sensing matrices $\mathbf{\Phi}$.]{
\includegraphics[width=0.45\textwidth]{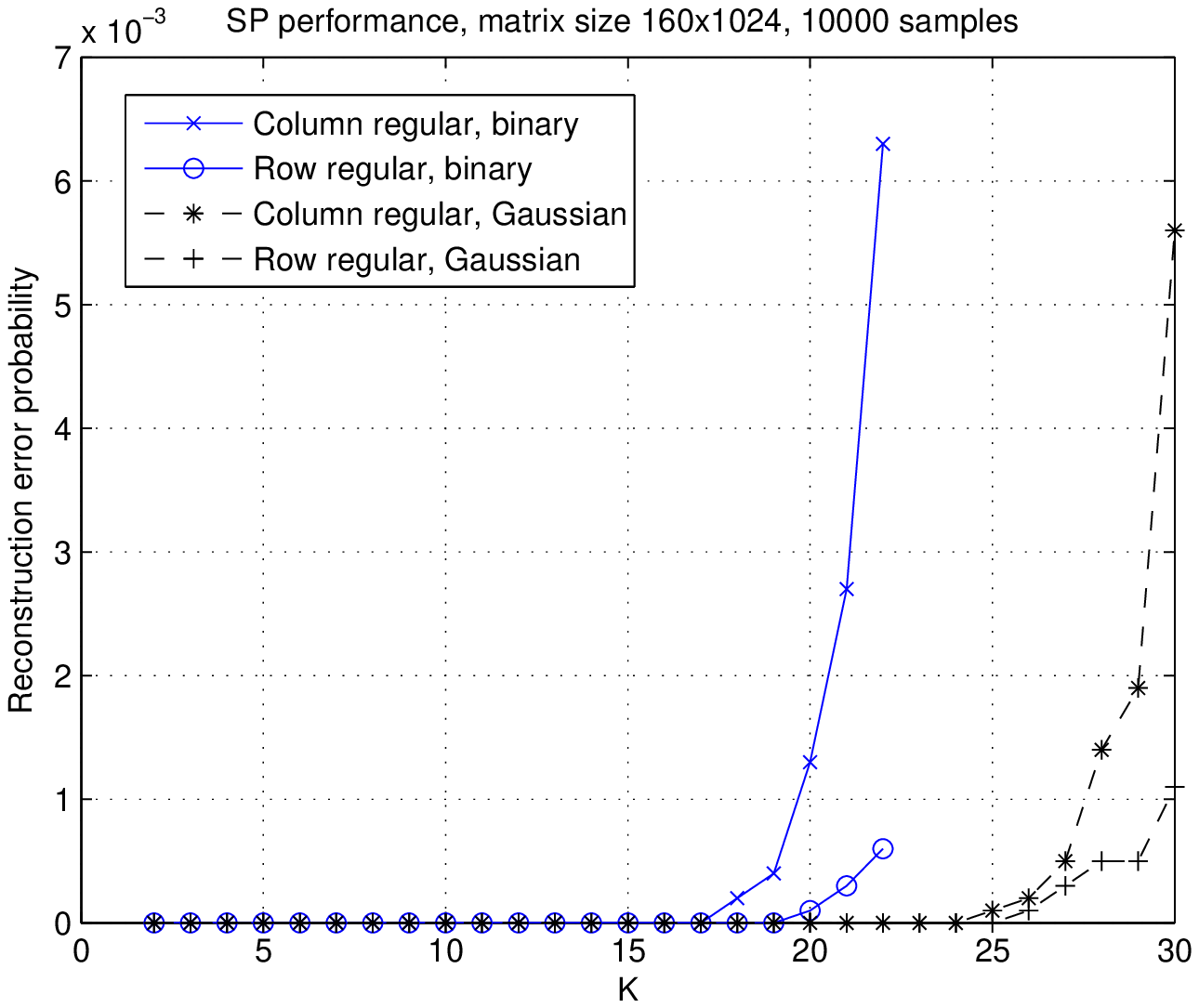}
\label{fig:Standard_SP}
}
\caption {\label{fig:Standard_OMP_SP}Reconstruction error probability of the standard OMP \subref{fig:Standard_OMP} and SP \subref{fig:Standard_SP} algorithms.}
\end{figure}

\begin{figure}
  \centering
  \includegraphics[width=0.5\textwidth]{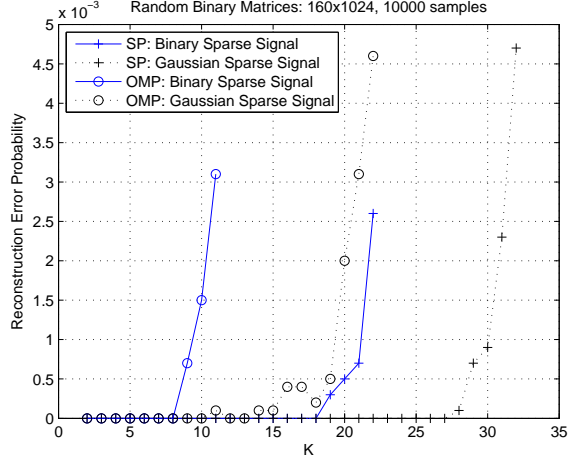}
  \caption{\label{fig:Standard_OMP_SP_Random}Performance of the OMP and SP algorithms with random sensing matrices $\mathbf{\Phi}$.}
\end{figure}

Figures~\ref{fig:Standard_OMP_SP} a) and b) illustrate the 
performance of the standard OMP and SP reconstruction algorithms applied to sensing matrices 
constructed using column-regular LDPC codes or row-regular LDPC codes. The performance plots are given for both binary and Gaussian sensing
vectors. Figure~\ref{fig:Standard_OMP_SP_Random} shows the same performance curves for a randomly generated sensing matrix of the same dimension, using realizations of i.i.d. Bernoulli random
variables.

The findings are rather interesting: they indicate that, at least with respect to the cut-off density (defined as the smallest value of $K$ for which the reconstruction error is not confined below some small
threshold probability) the LDPC code based sensing exhibit almost no performance loss compared to the random like matrices. For example, the cut-off for binary sensing vectors under OMP and
SP reconstruction for random-like sensing matrices are equal to $8$ and $18$, respectively. For Gaussian signals, these thresholds change to $12$ and $27$, respectively. For LDPC code based matrices, 
the cut-off thresholds are only slightly smaller: $7$ and $19$, and $11$ and $24$, respectively. Note that row-regular codes seem to slightly outperform column-regular codes for this particular example,
although we did not notice a general trend in performance gain/loss that can be attributed to this particular code characteristic - sometimes column-regular codes perform better than row-regular codes,
and sometimes the opposite is true.  

Next, we present simulation results for different BP correlation maximization strategies, coupled with the OMP and SP algorithm.
The simulation results show that non-systematic LDPC code sensing matrices do not perform well under CS reconstruction when $K$ is small. 
However, their systematic forms -- forms given by $\mathbf{H}_s=\left[\mathbf{I}| \mathbf{P}\right]$, where $\mathbf{I}$ denotes the identity matrix of size $(m-s)$ -- of the matrices have very good performance. 
This suggests that in the high interference regime, systematic parity-check matrices work better than non-systematic ones. This may confirm the recently described results of~\cite{calderbank1},
which can be very roughly stated as follows: for very low SNR regimes, ``bad'' LDPC codes outperform ``good'' ones.

\begin{figure}[ht]
\centering
\subfigure[Simulation results for the SP algorithm using the biased list-based BP decoder and the MMPC algorithm as its correlation maximization step.]{
\includegraphics[width=0.44\textwidth]{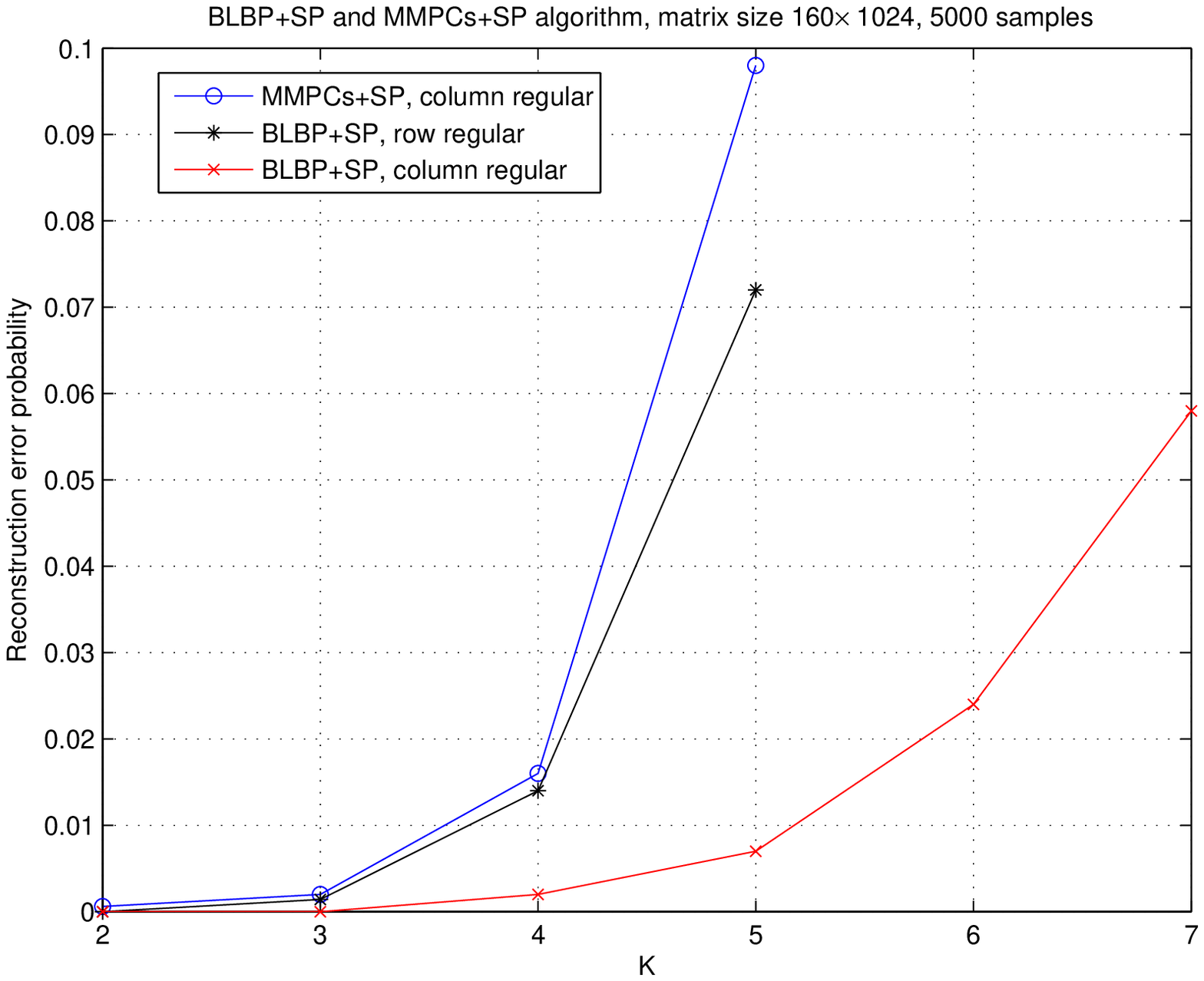}
\label{fig:List-Decoder}
}
\subfigure[Reconstruction error probability for the OMP algorithm using MBBP as its correlation maximization step.]{
\includegraphics[width=0.42\textwidth]{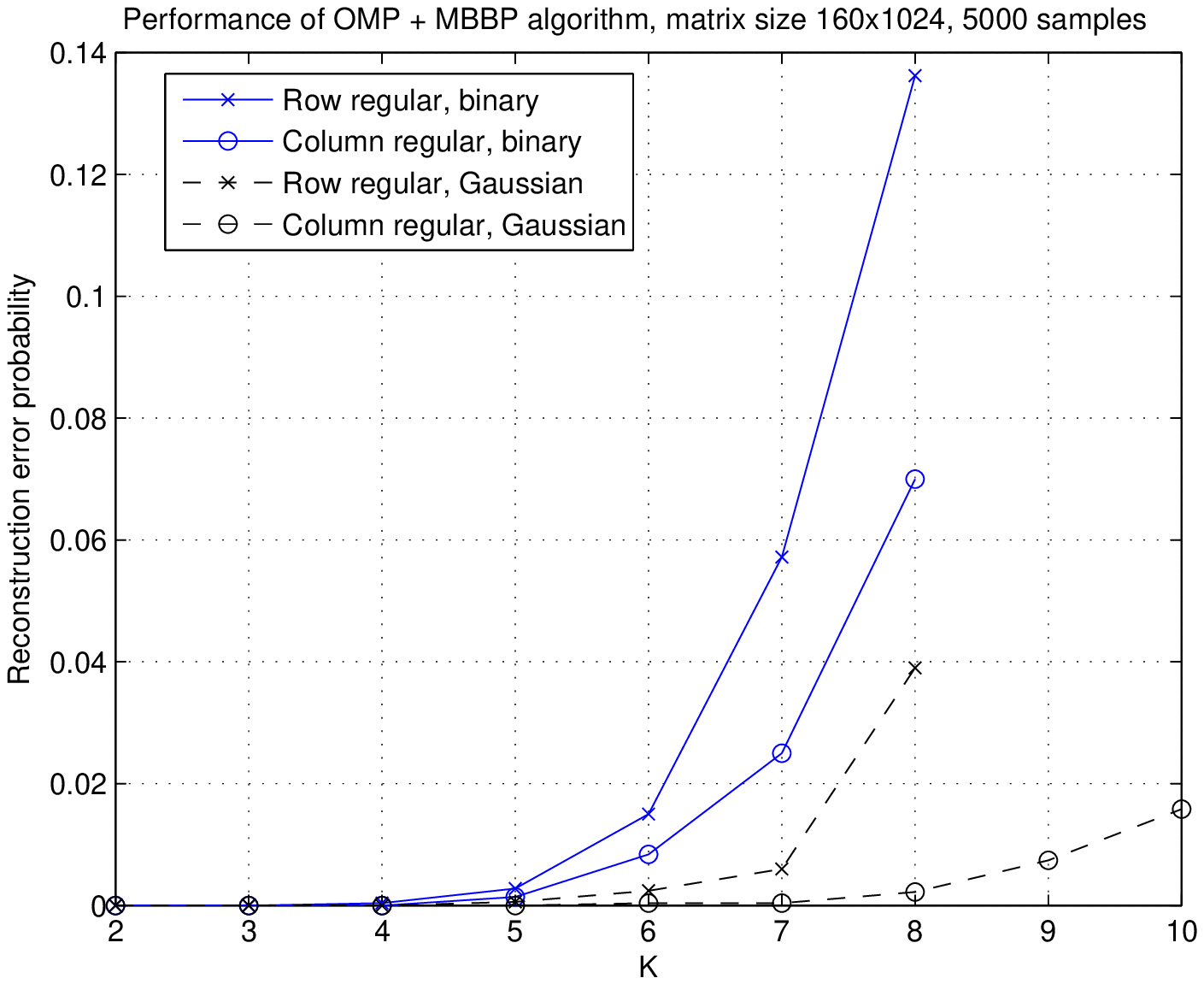}
\label{fig:MBBP_Gaussian}
}
\caption {\label{fig:TT}Performance of list-decoding algorithms coupled with OMP and SP.}
\end{figure}

Figures~\ref{fig:List-Decoder} and~\ref{fig:MBBP_Gaussian} illustrate 
the performance of the SP algorithm with a BP correlation maximization component and of the MMPCs-OMP algorithm. In both case, we used systematic forms of 
row-regular and column-regular parity-check matrices, resulting in sensing matrices of size $160\times 1024$. For the BP-SP algorithm, we used a list of size $L=20$. 
The reconstruction error probability is defined as the probability of not finding all the correct columns in the output list. OMP algorithms combined with BP methods for identifying 
codewords with largest correlations exhibit slow convergence, which does not make them amenable for CS applications. 

Figure \ref{fig:MBBP_Gaussian} shows the simulation results for the OMP algorithm using MBBP as its correlation maximization step 
for both binary and Gaussian signals.  We used four parity-check matrices ($p=4$): one of them is either row-regular matrix or column-regular, 
while the remaining three parity-check matrices represent systematic forms of the original matrix, obtained via Gaussian elimination applied with different column orders. 
The reconstruction error probability is once again defined as the probability of not finding all the correct columns in the 
output list. As can be seen, the MBBP method performs well for Gaussian signals \textbf{x}, although binary vectors still represent a reconstruction challenge.

\begin{figure}
  \centering
  \includegraphics[width=0.5\textwidth]{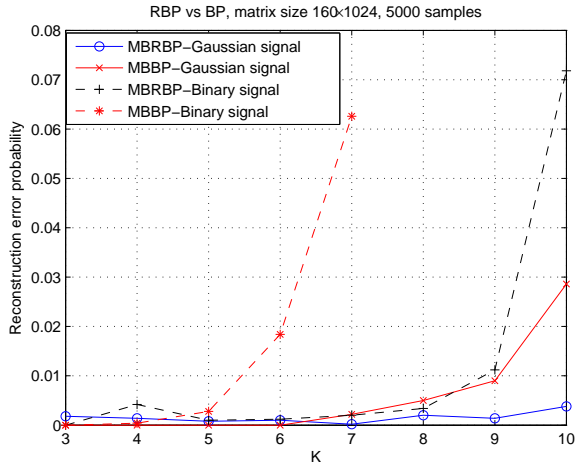}
  \caption{\label{fig:RBP_binary}Reconstruction error probability of the OMP algorithm for binary and Gaussian sensing vectors, using BP or RBP  as its correlation maximization step.}
\end{figure}

Figure \ref{fig:RBP_binary} compares the performance of the MBBP algorithm when using RBP (MBRBP) instead of the standard BP algorithm (MBBP). The parity-check matrix used for these 
simulations is column-regular. For the simulations, the parameters were set to $\psi_0=0.8$ and $\psi_1=0.99$. 
One can see that for small values of $K$, the standard BP algorithm performs slightly better than the RBP algorithm. However, RBP performs much better than BP for larger values of $K$.

\begin{figure}[ht]
\centering
\subfigure[Performance of the OMP and SP algorithm for  ``short'' LDPC codes based sensing matrices.]{
\includegraphics[width=0.45\textwidth]{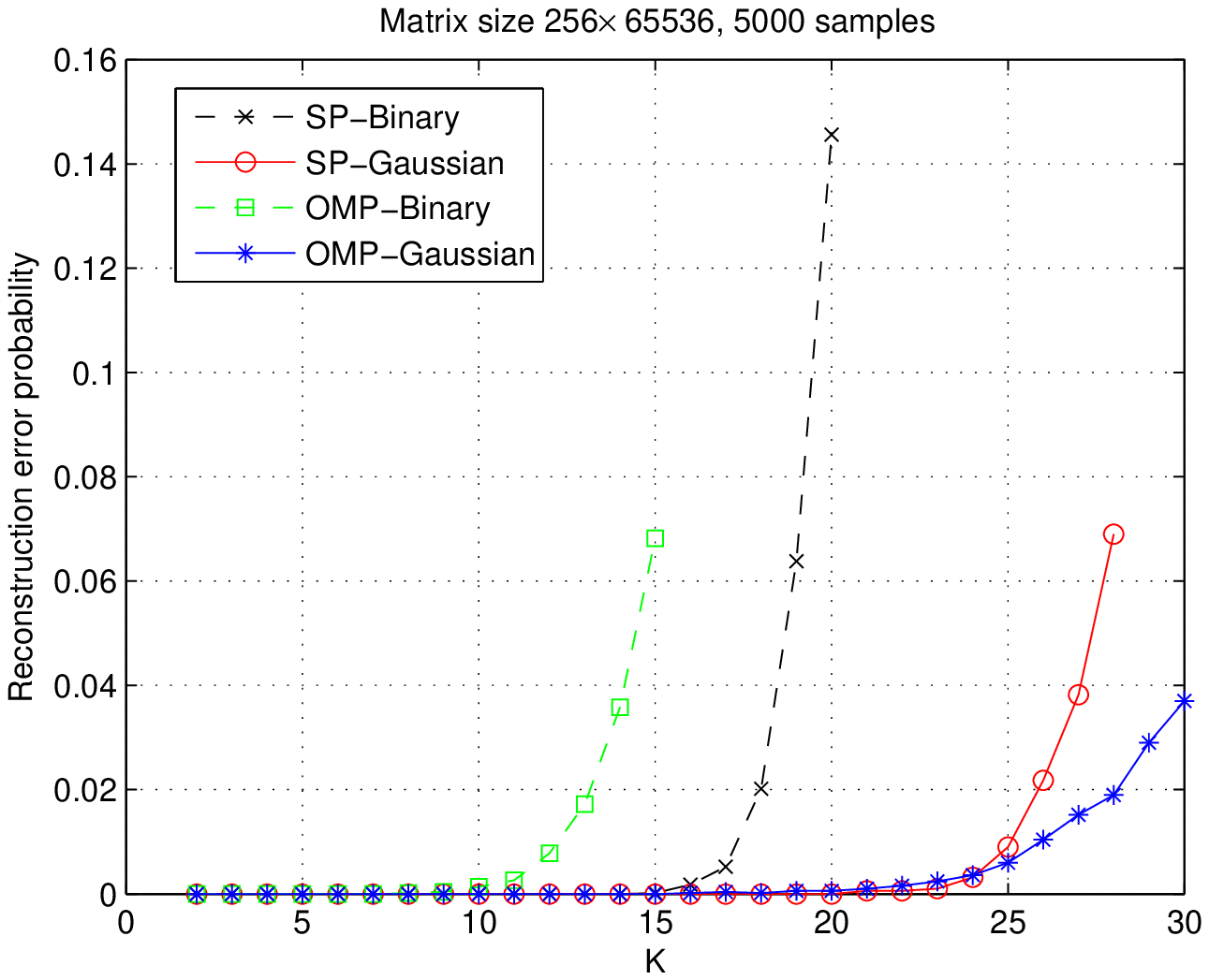}
\label{fig:Standard-Long}
}
\subfigure[Performance of the OMP and SP algorithm with list decoding BP methods for ``tall'' LDPC codes based sensing matrices.]{
\includegraphics[width=0.45\textwidth]{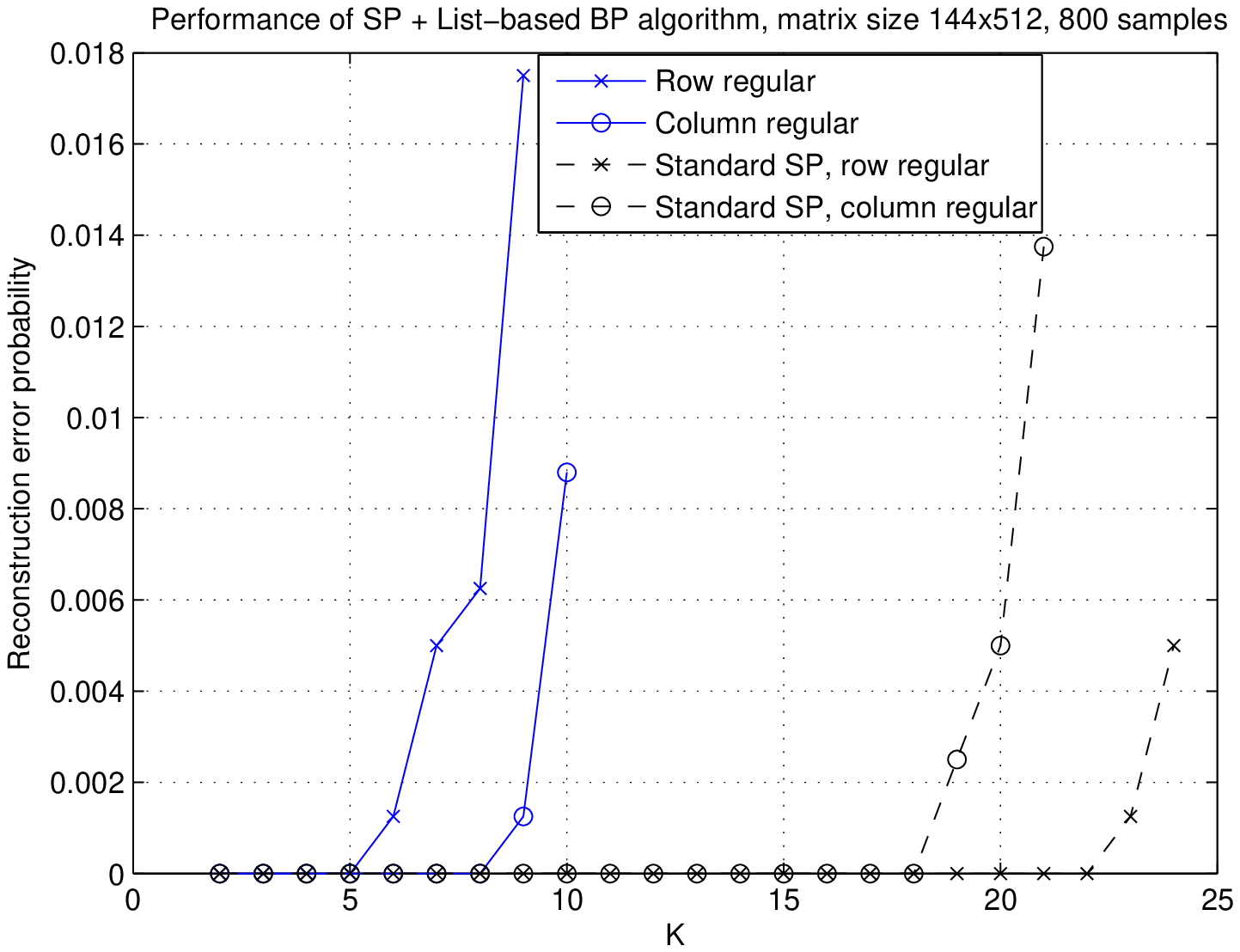}
\label{fig:List-Shorter}
}
\caption {\label{fig:T}Performance of the OMP/SP algorithms on large and small structured sensing matrices.}
\end{figure}

Finally, we present simulation results for codelengths that differ from $160$. Figure~\ref{fig:Standard-Long} illustrates the performance of a large dimensional
sensing matrix ($256 \times 65536$) constructed using LDPC codes. Figure~\ref{fig:List-Shorter} illustrates the influence of the choice of column- and row-regular matrices on the performance
of the sensing scheme for a parity-check matrix based on a length $144$ LDPC code. In Figure~\ref{fig:List-Shorter}, the reconstruction algorithm used is the list-decoding SP method.

\section{Conclusions} \label{sec:conclusion}

We described a new method for structured design of compressive sensing matrices based on LDPC codes. The special structure of the matrices makes them amenable
for low-complexity reconstructions of supersparse signals via new variants of BP-based decoding. We presented both theoretical results for the bounds on the coherence
parameter and the RIP of such matrices, as well as simulation results for the BP based reconstruction methods.

\subsection{\label{sub:Proof-of-random-Bernoulli}Proof of Proposition \ref{prop:perf-random-Bernoulli}}

To prove Proposition \ref{prop:perf-random-Bernoulli}, we need to
prove two claims.
\begin{enumerate}
\item When $K,m,N$ are sufficiently large with $m \geq c\, K^{2}\log N$,
$\underset{i\ne j}{\max}\frac{1}{m}\left|\left\langle \bm{x}_{i},\bm{x}_{j}\right\rangle \right|\le\frac{1}{2K}$
with overwhelming probability. 
\item Conversely, for any $\epsilon>0$, if $K,m,N$ are large enough with
$m=O\left(K^{2-\epsilon}\log N\right)$, $\underset{i\ne j}{\max}\frac{1}{m}\left|\left\langle \bm{x}_{i},\bm{x}_{j}\right\rangle \right|>\frac{1}{2K}$
with probability at least $1/2$.
\end{enumerate}
\vspace{0cm}

To prove the first part, we use large deviations technique and the union
bound. Let us first fix $i\ne j$. Note that $\frac{1}{m}\left\langle \bm{x}_{i},\bm{x}_{j}\right\rangle =\frac{1}{m}\sum_{k=1}^{m}X_{i,k}X_{j,k}$
where $X_{i,k}$ and $X_{j,k}$ are the $k^{th}$ elements of $\bm{x}_{i}$
and $\bm{x}_{j}$, respectively. Let $Z_{k}=X_{i,k}X_{j,k}$. Clearly,
$Z_{k}$'s, $k=1,2,\cdots,m$, are independent Bernoulli random variables.
The generic moment generating function of such variables is given by 
\[
\mbox{E}\left[e^{\theta Z}\right]=\frac{1}{2}e^{\theta}+\frac{1}{2}e^{-\theta}.
\]
According to the large deviations principle~\cite{dembo}, one has
\begin{equation}
\underset{m\rightarrow\infty}{\lim}\frac{1}{m}\log\Pr\left(\underset{i\ne j}{\max}\frac{1}{m}\left|\left\langle \bm{x}_{i},\bm{x}_{j}\right\rangle \right|>x\right)=-I\left(x\right),\;\forall x>0,\label{eq:Large-deviations}
\end{equation}
where $I\left(x\right)$ is the rate function, given by
\begin{align}
I\left(x\right) & =\underset{\theta}{\max}\; x\theta-\log\left(\mbox{E}\left[e^{\theta Z}\right]\right)\nonumber \\
 & =\underset{\theta}{\max}\; x\theta-\log\left(\frac{1}{2}e^{\theta}+\frac{1}{2}e^{-\theta}\right).\label{eq:I-x}
\end{align}
We compute the derivative of $x\theta-\log\left(\frac{1}{2}e^{\theta}+\frac{1}{2}e^{-\theta}\right)$
with respect to $\theta$ and set it to zero. The optimal $\theta$
needed to achieve the maximum in the definition of the rate function $I\left(x\right)$ is given by 
\begin{equation}
\theta^{*}=\frac{1}{2}\log\frac{1+x}{1-x}.\label{eq:optimal-theta}
\end{equation}
Substituting the explicit form for $\theta^{*}$ into the definition of $I\left(x\right)$ gives
\begin{equation}
I\left(x\right)=\frac{1}{2}x^{2}+o\left(x^{2}\right).\label{eq:rate-function}
\end{equation}
Hence, according to the large deviations principle in (\ref{eq:Large-deviations}), and using $o_v(1)$ to denote a value vanishing with $v$,
\begin{align*}
 & \Pr\left(\underset{i\ne j}{\max}\frac{1}{m}\left|\left\langle \bm{x}_{i},\bm{x}_{j}\right\rangle \right|>x\right)\\
 & =\exp\left[-m\left(\frac{1}{2}x^{2}+o\left(x^{2}\right)\right)\left(1+o_{m}\left(1\right)\right)\right]\\
 & \le\exp\left[-2m\left(\frac{1}{2}x^{2}+o\left(x^{2}\right)\right)\right]
\end{align*}
when $m$ is sufficiently large. We now apply the union bound for all
possible $1\le i\ne j\le N$. One has 
\begin{align}
 & \Pr\left(\underset{i\ne j}{\max}\frac{1}{m}\left|\left\langle \bm{x}_{i},\bm{x}_{j}\right\rangle \right|>\frac{1}{2K}\right)\nonumber \\
 & \le{N \choose 2}\exp\left[-2m\left(\frac{1}{2}x^{2}+o\left(x^{2}\right)\right)\right]\nonumber \\
 & \le\exp\left[-mx^{2}\left(1+o_{x}\left(1\right)\right)+2\log N\right].\label{eq:union-bound}
\end{align}
Set $x=\frac{1}{2K}$. Clearly, as $K,m$ are sufficiently large
and $m\ge c_{1}K^{2}\log N$ for some constant $c_{1}$, the probability
in (\ref{eq:union-bound}) can be made as small as $e^{-mc_{2}}$
for some constant $c_{2}$. This proves the first part of the claim. 

The converse is proved as follows. Note that a lower bound for $\Pr\left(\underset{i\ne j}{\max}\frac{1}{m}\left|\left\langle \bm{x}_{i},\bm{x}_{j}\right\rangle \right|>\frac{1}{2K}\right)$
is obtained by fixing $i=1$, i.e., 
\[
\Pr\left(\underset{i\ne j}{\max}\frac{1}{m}\left|\left\langle \bm{x}_{i},\bm{x}_{j}\right\rangle \right|>\frac{1}{2K}\right)\ge\Pr\left(\underset{j\ne1}{\max}\frac{1}{m}\left|\left\langle \bm{x}_{1},\bm{x}_{j}\right\rangle \right|>\frac{1}{2K}\right).
\]
It suffices to prove the lower bound is nontrivial. Clearly, $\frac{1}{m}\left\langle \bm{x}_{1},\bm{x}_{j}\right\rangle =\frac{1}{m}\sum_{k=1}^{m}X_{1,k}X_{j,k}=\frac{1}{m}\sum_{k=1}^{m}Z_{k}$
where $Z_{k}$'s are independent Bernoulli random variables with parameter
$p=1/2$. The large deviations analysis in (\ref{eq:Large-deviations})
and (\ref{eq:rate-function}) is still valid. More importantly, the variables $\frac{1}{m}\left\langle \bm{x}_{1},\bm{x}_{j}\right\rangle $'s
are independent for different values of $j$. Hence, 
\begin{align*}
\Pr\left(\underset{j\ne1}{\max}\frac{1}{m}\left|\left\langle \bm{x}_{1},\bm{x}_{j}\right\rangle \right|>\frac{1}{2K}\right) & =1-\Pr\left(\underset{j\ne1}{\max}\frac{1}{m}\left|\left\langle \bm{x}_{1},\bm{x}_{j}\right\rangle \right|\le\frac{1}{2K}\right)\\
 & =1-\Pr\,^{N-1}\left(\frac{1}{m}\left|\left\langle \bm{x}_{1},\bm{x}_{j}\right\rangle \right|\le\frac{1}{2K}\right).
\end{align*}
It suffices to prove 
\begin{equation}
\Pr\,^{N-1}\left(\frac{1}{m}\left|\left\langle \bm{x}_{1},\bm{x}_{j}\right\rangle \right|\le\frac{1}{2K}\right)\le\frac{1}{2}.\label{eq:converse-goal}
\end{equation}

To simplify the notation, let $P_{\mbox{goal}}$ denote $\Pr\,^{N-1}\left(\frac{1}{m}\left|\left\langle \bm{x}_{1},\bm{x}_{j}\right\rangle \right|\le\frac{1}{2K}\right)$
and let $\bar{Z}$ denote the random variable $\frac{1}{m}\left|\left\langle \bm{x}_{1},\bm{x}_{j}\right\rangle \right|$.
Note that 
\begin{align*}
\log P_{\mbox{goal}} & =\left(N-1\right)\log\Pr\left(\bar{Z}\le\frac{1}{2K}\right)\\
 & =\left(N-1\right)\log\left(1-\Pr\left(\bar{Z}>\frac{1}{2K}\right)\right).
\end{align*}
When $m$ is sufficiently large, $\Pr\left(\bar{Z}>\frac{1}{2K}\right)$
is sufficiently small according to the large deviations principle.
Hence, for sufficient large $K$ and $m$, one has 
\begin{align*}
\log P_{\mbox{goal}} & =-\left(N-1\right)\Pr\left(\bar{Z}>\frac{1}{2K}\right)\left(1+o_{m}\left(1\right)\right)\\
 & \le-\frac{N}{2}\Pr\left(\bar{Z}>\frac{1}{2K}\right)\\
 & =-\frac{N}{2}\exp\left(-mI\left(\frac{1}{2K}\right)\left(1+o_{m}\left(1\right)\right)\right)\\
 & \le-\frac{N}{2}\exp\left(-2mI\left(\frac{1}{2K}\right)\right)\\
 & \overset{\left(\ref{eq:rate-function}\right)}{=}-\frac{N}{2}\exp\left(-2m\frac{1}{8K^{2}}\left(1+o_{K}\left(1\right)\right)\right)\\
 & \le-\frac{1}{2}\exp\left(-\frac{m}{2K^{2}}+\log N\right).
\end{align*}
Suppose that $m=cK^{2-\epsilon}\log N$ for some constant $c$.
\begin{align*}
\log P_{\mbox{goal}} & \le-\frac{1}{2}\exp\left(-\frac{c}{2}\frac{\log N}{K^{\epsilon}}+\log N\right).
\end{align*}
 No matter how large $c$ is, as long as $K,m,N$ are sufficiently
large, we have $c/K^{\epsilon}<1$. Hence, 
\begin{align*}
\log P_{\mbox{goal}} & \le-\frac{1}{2}\exp\left(\frac{1}{2}\log N\right)\le-\frac{1}{2}\sqrt{N}\le-\log2
\end{align*}
for large $K,m,N$. The desired (\ref{eq:converse-goal}) therefore holds. This completes the proof. 

\section*{Acknowledgements}
The authors are grateful to Rudiger Urbanke and Pascal Vontobel for useful discussions. Parts of the results were presented at ISIT 2009, Seoul, Korea.
\bibliographystyle{IEEEtran}
\bibliography{CompressedSensing01,LDPC_Group_Bibfile}
\end{document}